\documentclass{aa}  

\usepackage{graphicx}
\usepackage{xcolor} 
\usepackage[varg]{txfonts}
\usepackage{rotating} 
\usepackage{url}  
\usepackage{mathtools, amsmath, amssymb, amsfonts} 
\usepackage{ccicons} 
\usepackage{mathabx}
\usepackage{multirow}
\usepackage{hyperref}
\begin{document}

   \title{Dust clearing by radial drift in evolving protoplanetary discs}

   \subtitle{}

   \author{Johan Appelgren
          \and Michiel Lambrechts
          \and Anders Johansen
          }

   \institute{Lund Observatory, Department of Astronomy and Theoretical Physics, Lund University, Box 43, 22100 Lund, Sweden \\
   email: johan@astro.lu.se
             }

   \date{Received ; accepted }

 
  \abstract
 {      Recent surveys have revealed that protoplanetary discs typically have dust masses that appear to be insufficient to account for the high occurrence rate of exoplanet systems. 
We demonstrate that this observed dust depletion is consistent with the radial drift of pebbles.
Using a Monte Carlo method we simulate the evolution of a cluster of protoplanetary discs using a 1D numerical method to viscously evolve each gas disc together with the radial drift of dust particles that have grown to 100 $\mu$m in size. 
For a 2 Myr-old cluster of stars, we find a slightly sublinear scaling between the gas disc mass and the gas accretion rate ($M_\mathrm{g} \propto \dot{M}^{0.9}$). 
However, for the dust mass we find that evolved dust discs have a much weaker scaling with the gas accretion rate, with the precise scaling depending on the age at which the cluster is sampled and the intrinsic age spread of the discs in the cluster.
Ultimately, we find that the dust mass present in protoplanetary discs is on the order of 10-100 $M_\Earth$ in 1-3 Myr-old star-forming regions, a factor of 10 to 100 depleted from the original dust budget. As the dust drains from the outer disc, pebbles pile up in the inner disc and locally increase the dust-to-gas ratio by  up to a factor of four above the initial value. In these regions of high dust-to-gas ratio we find conditions that are favourable for planetesimal formation via the streaming instability and subsequent growth by pebble accretion.
We also find the following scaling relations with stellar mass within a 1-2 Myr-old cluster: a slightly super-linear scaling between the gas accretion rate and stellar mass ($\dot{M} \propto M_\star^{1.4}$), a slightly super-linear scaling between the gas disc mass and the stellar mass  ($M_\mathrm{g} \propto M_\star^{1.4}$), and a super-linear relation between the dust disc mass and stellar mass ($ M_\mathrm{d} \propto M_\star^{1.4-4.1}$).
}


   \maketitle
%

\section{Introduction}
   
Observations of star-forming regions have now provided a sufficiently large sample of protoplanetary discs to allow statistical analyses of their properties \citep{2016ApJ...828...46A,2017A&A...604A.127M,2017A&A...600A..20A,2018ApJ...869L..41A}. Disc properties such as gas and dust disc masses and their respective outer radii are important because they directly influence the formation of planets by setting the available mass reservoir and the time-span in which they can form. However, it is not trivial to interpret the distribution of disc properties in star-forming regions. For instance,  the mass reservoir in observed protoplanetary discs does not appear to be large enough to be able to form the commonly observed exoplanetary systems \citep{2014MNRAS.445.3315N,2018A&A...618L...3M}.  

Star formation takes place in giant molecular clouds. These clouds fragment into dense cores that in turn collapse gravitationally to form stars \citep{2007ARA&A..45..565M}. Conservation of angular momentum causes some of the gas and dust in the cloud core to land in a disc surrounding the protostar. The young protostar grows in mass by accreting gas from the surrounding disc. This accretion is driven by a redistribution of angular momentum in the disc, although the physical process behind the angular momentum redistribution is uncertain \citep{2014prpl.conf..411T}. 

In order to model the evolution of the dust component in protoplanetary discs, detailed models have been developed that include \cite{2005ApJ...623..482T,2005ApJ...627..286T}, \cite{2008A&A...480..859B}, and \cite{2010A&A...513A..79B,2012A&A...539A.148B}.  
These works show that compact millimetre(mm)-sized dust particles are drained in less than 1 Myr, but that fluffy grains are able to remain for longer. \cite{2005ApJ...627..286T} also found that in the presence of strong photoevaporation, the inner gas disc can be cleared out before the dust drains. The dust remaining in the outer disc then piles up near the inner edge of the gas disc. \cite{2005ApJ...623..482T} also showed that the rapid radial drift of dust weakens the emission from the disc at mm wavelengths, and that the dust needs to be in the size range between 1 mm and 1 cm for the emission at mm wavelengths to remain present for more than 1 Myr. 
 
Surveys of protoplanetary discs provide insight into scaling relations between disc properties and the stellar mass. A super-linear scaling relation between the stellar mass and the dust disc mass  was observed by \cite{2016ApJ...831..125P}. In the gas accretion-rate-versus-stellar-mass scaling relation, indications of a broken power have been found \citep{2017A&A...604A.127M,2017A&A...600A..20A}.  \cite{2017A&A...604A.127M} found $\dot{M}\propto M_\star^{4.31}$ for $M_\star \leq 0.3 \ M_\sun$ and $\dot{M}\propto M_\star^{0.56}$ for $M_\star \geq 0.3 \ M_\sun$, and \cite{2017A&A...600A..20A} found $\dot{M}\propto M_\star^{4.58}$ for $M_\star \leq 0.2 \ M_\sun$ and $\dot{M}\propto M_\star^{1.37}$ for $M_\star \geq 0.2 \ M_\sun$.  \cite{2017A&A...604A.127M} also made a single power-law fit and found $\dot{M}\propto M_\star^{2.3}$.

Some progress has been made in trying to understand the scaling relations mentioned above. \cite{2006ApJ...645L..69D} used a disc formation and evolution model to investigate the scaling between the gas accretion rate and the stellar mass. These latter authors found $\dot{M} \propto M_\star^{1.8}$, if the ratio between the rotational and gravitational energy of the cloud core and the viscous $\alpha$ parameter are kept the same for cloud cores of different masses. This agrees well with observations measuring the gas accretion rate and stellar mass scaling with a single power law \citep{2003ApJ...592..266M,2004A&A...424..603N,2016A&A...585A.136M,2017A&A...600A..20A}, but is in contrast with recent indications of a broken power law scaling \citep{2017A&A...604A.127M,2017A&A...600A..20A}.

Using Monte Carlo methods, \cite{2009ApJ...704..989A}  modelled viscous disc evolution around 1000 protoplanetary discs including photoevaporation and gap opening due to the presence of a giant planet. With a maximum initial gas disc mass of$\approx\!  0.12$ $M_\sun$ and a mean of $ 0.0045$ $M_\sun$, the authors find median gas accretion rates between $\simeq 0.5\times 10^{-8}$ $M_\sun$ yr$^{-1}$ at $t \simeq 0.1$ Myr and $\simeq 10^{-10}$ $M_\sun$ y$r^{-1}$ at $t \simeq 4$ Myr. Disc lifetimes range between 2.3 and 10.7 Myr.

 In contrast to these studies we aim to model the formation and evolution of a representative sample of protoplanetary discs in a star-forming region by evolving both the gas and dust component. We are especially interested in the evolution of the dust mass, as it sets the initial reservoir of mass that is available for planet formation and because it is an important observable in disc surveys. 

This paper explores the formation and evolution of protoplanetary gas and dust discs. In Sect. \ref{sec:model} we describe the numerical model that we use for the formation and evolution of the protoplanetary disc. We do not use a
singular istothermal sphere \citep{1977ApJ...214..488S,2005A&A...442..703H} to calculate the infall of mass onto the disc from the collapsing core, but instead a Bonnor-Ebert sphere \citep{1956MNRAS.116..351B,1957ZA.....42..263E} to model the cloud core, as in \cite{2013ApJ...770...71T}. 
 We combine the infall with viscous evolution and disc gravitational instability in a one-dimensional numerical model to simulate disc evolution. At first, we look at the formation and evolution of a single disc and then we explore the effects of the angular momentum budget of the cloud core on disc evolution.  We then use a Monte Carlo method to study a cluster of discs in a star-forming region that evolves over several million years. 
 In Sect. \ref{sec:results}, we present our main results. We find that during evolution of the disc the dust component goes through three main phases: the formation phase, a phase of slow radial dust drift due to high gas surface densities, and one of rapid radial dust drift once the gas surface density drops sufficiently. Our most important conclusion is that the low observed dust disc masses in star-forming regions (\cite{2016ApJ...828...46A} found $\overline{M}_\mathrm{d} \approx\! 15\ M_\Earth$ for 1-3 Myr-old star-forming regions) is consistent with radial drift of dust pebbles.
 In Sect. \ref{sec:scaling} we focus on the observationally relevant scaling relationships. Of particular interest is the relation between the main observables: the stellar gas accretion rate and the dust disc mass.
   In Sect. \ref{sec:discussion} we discuss the results and compare them to observations and examine their implications on planet formation. Finally, in Sect. \ref{sec:conclusions} we summarise our conclusions.

\section{Model}
\label{sec:model}

\subsection{Disc formation}

We use a model for the collapse of the cloud core and the formation of the circumstellar disc similar to the one described in \cite{2013ApJ...770...71T}. The molecular cloud core is modelled as a Bonnor-Ebert sphere \citep{1956MNRAS.116..351B,1957ZA.....42..263E} with a constant temperature $T_\mathrm{cc} = 10$ K throughout the entire cloud and a solid-body rotation rate $\Omega_0$. In order to trigger the gravitational collapse of the cloud core, we follow \cite{2013ApJ...770...71T} and increase its density by a factor $f=1.4$. For a discussion on the effect of using a different value of $f$ we refer to \cite{2013ApJ...770...71T}.

The cloud core collapses inside out, with the radius of the collapse front, $r_\mathrm{cf}$, expanding outwards in time. \cite{2013ApJ...770...71T} gives the time it takes for the collapse to reach a radius $r_\mathrm{cf}$ in the cloud core as
\begin{align}
t_\mathrm{collapse} =\sqrt{ \dfrac{r_\mathrm{cf}^3}{2GM(r_\mathrm{cf})}}\int_0^1\dfrac{\mathrm{d}R}{\sqrt{\dfrac{1}{f}\ln R + \dfrac{1}{R} - 1}},
\end{align}
where $G$ is the gravitational constant, $M(r_\mathrm{cf})$ is the mass inside the collapse front radius, and $R = r_\mathrm{cf}/r_\mathrm{cc}$, where $r_\mathrm{cc}$ is the radius of the cloud core.

The rate at which gas will fall onto the disc and star, $\dot{M}_\mathrm{infall}$, is given by 
\begin{align}
        \dot{M}_\mathrm{infall}(t) = 4\pi r_\mathrm{cf}(t)^2\rho(r_\mathrm{cf}(t))\dfrac{\mathrm{d}r_   \mathrm{cf}}{\mathrm{d}t},
\end{align}
where $\rho(r_\mathrm{cf})$ is the density in the cloud core at $r_\mathrm{cf}$ and $\mathrm{d}r_\mathrm{cf}/\mathrm{d}t$ is the velocity at which the collapse front expands outwards. A derivation for an expression for $\mathrm{d}r_\mathrm{cf}/\mathrm{d}t$ can be found in Appendix \ref{app:collapse}.

 The gravitational collapse of the cloud conserves angular momentum. This causes the infalling material to land on an orbit in the protoplanetary disc where its Keplerian angular momentum is the same as the angular momentum it had in the cloud core. The  outermost radius at which material lands at a given time is called the centrifugal radius, $a_\mathrm{c}$, and is given by
\begin{align}
a_\mathrm{c}(t) = \dfrac{\Omega_0^2 r_\mathrm{cf}(t)^4}{G M(r_\mathrm{cf}(t))}. \label{eq:ac}
\end{align}

The infalling material is spread out across the disc within the centrifugal radius such that its angular momentum is conserved. The change in the surface density of the disc is given by the following equation from \cite{2005A&A...442..703H}:
\begin{align}
S_\mathrm{g}(r,t) = \dfrac{\dot{M}_\mathrm{infall}}{8\pi a_\mathrm{c}^2}\left(\dfrac{r}{a_\mathrm{c}}\right)^{-3/2}\left[1-\left(\dfrac{r}{a_\mathrm{c}}\right)^{1/2}\right]^{-1/2}, \label{eq:SourceTerm}
\end{align}
where $r$ is the orbital radius in the disc. For the dust component of the infall we assume the source term to be
\begin{align}
S_\mathrm{d}(r,t) = 0.01 S_\mathrm{g}(r,t),
\end{align}
where we assumed the dust-to-gas ratio to be 1/100 to match interstellar medium values and to maintain a fixed initial metallicity in the disc.

\subsection{Disc evolution}

\subsubsection{Gas disc}

We evolve the protoplanetary gas disc by viscous evolution, where the disc expands outwards as matter is accreted onto the central star. The viscous evolution of \cite{1981ARA&A..19..137P} is combined with the source term from Eq. \eqref{eq:SourceTerm}, which gives a total gas surface density evolution of
\begin{align}
        \dfrac{\partial\Sigma_\mathrm{g}}{\partial t} = \dfrac{3}{r}\dfrac{\partial}                    {\partial r}\left[r^{1/2}\dfrac{\partial}{\partial r}\left(\nu\Sigma_\mathrm{g}           r^{1/2}\right)\right] + S_\mathrm{g}(r,t),  \label{eq:pring}
\end{align}
 where $\Sigma_\mathrm{g}$ is the gas surface density and $\nu$ is the turbulent viscosity. To model the turbulent viscosity we use the standard $\alpha$ parametrisation, 
\begin{align}
        \nu = \alpha c_\mathrm{s}H_\mathrm{g};
\end{align}
\citep{1973A&A....24..337S}. Here, $c_\mathrm{s}$ is the sound speed in the disc and $H_\mathrm{g} = c_\mathrm{s}/\Omega_k$ is the scale height of the gas disc.
We assume that the total viscosity parameter is given by 
\begin{align}
        \alpha =  \alpha_\mathrm{min} +  \alpha_\mathrm{Q}  \label{eq:alpha},
\end{align} 
where $\alpha_\mathrm{min} = 10^{-2}$ is the minimum value throughout the disc. This model parameter was chosen pragmatically to evolve the $\alpha$-disc within the observed disc lifetimes of a few million years \citep{2001ApJ...553L.153H,2010A&A...510A..72F}. However, recent numerical efforts have shown that disc midplanes are almost laminar \citep{2013ApJ...769...76B,2015ApJ...801...84G} and point towards very low midplane alpha values (on the order of $\alpha_\mathrm{mid} \approx 10^{-4}$), in apparent agreement with protoplanetary disc scale heights \citep{2016ApJ...816...25P}.

 The effects of using a lower $\alpha_\mathrm{min}$ are discussed in Appendix \ref{app:alpha}. The parameter $\alpha_\mathrm{Q}$ models the effect of gravitational instabilities. When the disc is young it can become very massive relative to the central star and become gravitationally unstable.  Gravitational instability gives rise to gravitational torques in the disc that act as an increased disc viscosity. We model the gravitational torque following \cite{2010ApJ...713.1143Z} and \cite{2013ApJ...770...71T}  and use 
\begin{align}
        \alpha_\mathrm{Q} = A\exp\left(-BQ^{4}\right), \label{eq:alpha_q}
\end{align}
with $A=1$ and $B=1$, and where $Q$ is the Toomre parameter given by

\begin{align}
Q = \dfrac{c_\mathrm{s}\Omega_\mathrm{k}}{\pi G \Sigma_\mathrm{g}}.
\end{align}

 These choices for $A$ and $B$ cause $\alpha_Q$ to become important relative to $\alpha_\mathrm{min}$ when $ Q \lesssim 1.4$. Three-dimensional
global simulations have shown this to be an appropriate limit for gravitational torques to become important
  \citep[e.g.][]{2006ApJ...651..517B}. The precise choice of A and B has been demonstrated to have no significant effect on the long-term gas disc evolution \citep{2013ApJ...770...71T}. 

\subsubsection{Dust disc}
\label{sec:dustevo}

We use a simplified model for the dust component of the disc. We assume that the dust is of constant size ($r_\mathrm{p}\! =\! 100$ $\mu$m) throughout the disc. There are a number of motivations behind this choice. First, observations of the spectral index in envelopes around forming stars show that dust particles already as large as 100 $\mu$m to 1 mm in size could be present \citep{2014A&A...567A..32M,2019A&A...623A.147A}. Likewise, observations of very young stars indicate the presence of mm-sized dust already in the formation phase of the disc \citep{2018NatAs...2..646H}.  Observations of polarised emission from protoplanetary discs also indicate that the maximum particle size is $\approx\! 100$ $\mu$m if compact and spherical dust particles are assumed \citep{2016ApJ...831L..12K,2018ApJ...864...81O}.  

From laboratory experiments, it also appears that particle growth by coagulation in protoplanetary discs is inefficient past roughly mm sizes \citep{2010A&A...513A..56G,2010A&A...513A..57Z}. This holds for both silicates and ice particles, because at temperatures below 180 K the surface energy of ice grains is similar to that of silicates \citep{2019ApJ...873...58M}.

 Hence, our assumption of dust growth to 100 $\mu$m should accurately capture the drift of dust particles in the outer regions of the protoplanetary disc. Another simplification is that we do not include feedback of dust on gas, which is appropriate for low dust-to-gas ratios.

The aerodynamic coupling of the dust to the gas can be measured with the dimensionless Stokes number. In the Epstein regime, when the dust particles are smaller than the mean free path of the gas, and assuming vertical hydrostatic equilibrium in the disc, the Stokes number is given by 
\begin{align}
        \tau_\mathrm{s} =       \dfrac{\sqrt{2\pi} \rho_\mathrm{p} \ r_\mathrm{p}}      {\Sigma_\mathrm{g}}, \label{eq:Epstein}
\end{align}
where $\rho_\mathrm{p}$ is the material density of the dust particles. We set $\rho_\mathrm{p} = 1.6$ g cm$^{-3}$ following \cite{2010A&A...513A..79B}. We also include the transition to the Stokes drag regime when the dust particles are larger than the mean free path of the disc, which can occur in the inner disc, as described in \cite{2010EAS....41..187Y}. 

The dust surface density is evolved by advecting it with the gas and by letting the dust drift due to gas drag. We ignore turbulent diffusion of dust as the midplane is assumed to be almost laminar. In the early stages of disc evolution the advection term dominates because the high gas surface densities result in low Stokes numbers. Eventually, the gas surface density drops and Stokes numbers increase. The dust begins to decouple from the gas and radial drift of dust becomes important. We model this as an additional advection term on the dust. To calculate the radial dust velocity we use the equation given by \cite{1986Icar...67..375N}, 
\begin{align}
        v_\mathrm{r,d} = -\dfrac{2\tau_\mathrm{s}}{1+\tau_\mathrm{s}^2}\eta v_\mathrm{k},
\end{align}
where $\eta$ is a measure of the radial gas pressure support and $v_\mathrm{k}$ is the Keplerian velocity. Here, $\eta$ is given by
\begin{align}
        \eta = -\dfrac{1}{2}\left(\dfrac{H_\mathrm{g}}{r}\right)^2\dfrac{\partial\ln P}{\partial \ln r},
\end{align}
where the last factor is the radial pressure gradient.

\subsection{Disc temperature}

For simplicity, we assume that the disc is locally isothermal, with the temperature given as in the Hayashi MMSN model \citep{1981PThPS..70...35H}, but with the minimum  value set by the cloud core temperature of $T_\mathrm{cc} = 10$ K,
\begin{align}
T(r) =\begin{cases}
 280\left(\dfrac{L_\star}{L_\sun}\right)^{1/4}\left(\dfrac{r}{\mathrm{au}}\right)^{-1/2}\ \mathrm{K}, \quad &\mathrm{if}\ T\quad >T_\mathrm{cc} \\
 T_\mathrm{cc}, \quad &\mathrm{else},
\end{cases}
\end{align}
where $L_\star$ is the luminosity of the central star. For low-mass stars that are a few million years old or less, the stellar luminosity scales with the stellar mass roughly as
\begin{align}
L_\star \propto M_\star^{1-2}; \label{eq:Ls-Ms}
\end{align} 
\citep{2002A&A...382..563B}. The stellar luminosity is therefore expected to increase as the star grows in mass over time. However, our model does not take into account the heating of the disc by the infalling material dissipating its energy in the disc. Therefore, in our model, evolving the stellar luminosity with the stellar mass results in an exceedingly cold disc during the initial stages of a discs life. Instead we assume the luminosity to be given by a star of the same mass as the cloud core after a few years of evolution, according to Eq. \eqref{eq:Ls-Ms}. For this study, we chose to use 
\begin{align}
\dfrac{L_\star}{L_\sun} \propto \left( \dfrac{M_\star}{M_\sun} \right)^{2}.
\end{align}

In our model we chose to ignore the effects of viscous heating, which is thought to contribute to the temperature structure of the disc at short orbital radii. However, \cite{2019ApJ...872...98M}  used simulations of protoplanetary discs under non-ideal MHD conditions to show that in the inner disc, where accretion might be driven by magnetised disc winds and heating is caused by Joule dissipation rather than viscous dissipation, heating of the disc midplane by accretion is weak and stellar irradiation dominates the disc heating. Only at radii of $\lesssim\! 0.5$ au can viscous heating become important. Outside of the viscously dominated region, stellar irradiation sets the disc temperature. Although, during the early phases of the disc when accretion rates are higher, and especially during accretion bursts, the region dominated by viscous heating can extend out to a few tens of astronomical units \citep{2016Natur.535..258C}. As we see below, in our model the radial drift of the dust occurs from the outside in. Therefore, the focus of our interest is the outer disc at radii of more than at least a few tens of astronomical units. Viscous heating can be neglected at these radii.

\subsection{Simulations}
\label{sec:sims}

In this section we describe the simulations that we run and explain the reasoning behind certain modelling choices.
A few parameters are the same for all simulations. These are the temperature of the cloud core, which we assume to be $T_\mathrm{cc} = 10$ K, the value of the minimum viscous $\alpha$ parameter $\alpha_\mathrm{min} = 10^{-2}$, and the size of the dust, which we assume to be $r_\mathrm{p} = 100$ $\mu$m throughout the disc. We refer to Appendix \ref{app:alpha} for the effect of using a lower $\alpha_\mathrm{min}$ value and Appendix \ref{app:dustsize} for the effect of using a larger dust size.

In short, the simulations that we perform are: (a) a reference case for understanding the evolution of a single disc, (b) a set of runs where we look at the effect of the centrifugal radius and cloud core mass on disc evolution and (c) a set of Monte Carlo runs to model the evolution of a star-forming cluster.

\subsubsection{Reference case}

The first simulation we run is the reference case, which we use to understand how a single protoplanetary disc forms and evolves. This case consists of a cloud core mass of $M_\mathrm{cc} = 1$ $M_\sun$. We set the angular momentum of the cloud core such that the final centrifugal radius is $a_\mathrm{c} = 50$ au. 

\begin{table}[]
\caption{Final centrifugal radii for the different cloud core masses investigated. The cases within parenthesis were included when exploring the effect of the cloud core mass and the centrifugal radius on disc evolution. In the Monte Carlo study, centrifugal radii as low as the two lowest values were excluded owing to their computational cost and the highest two centrifugal radii were excluded because they give excessively long disc lifetimes.}
\label{tab:MassRcgrid}
\centering
\begin{tabular}{ll}
\hline\hline
$M_\mathrm{cc}$ {[}$M_\sun${]} & $a_\mathrm{c,fin}$ {[}au{]}   \\ \hline
0.1                               & (0.8), (1.6), 3.2, 6.4, 12.8 \\
0.5                               &  4, 8, 16, 32, 64             \\
1.0                               &  8, 16, 32, 64, 128           \\
2.0                               &  16, 32, 64, (128), (256)     \\
\hline   
\end{tabular}
\end{table}

\subsubsection{Changing the centrifugal radius and cloud core mass}

After we have analysed the reference case we examine the effect of changing the final centrifugal radius of the disc. For this we use the centrifugal radii listed for the 1 $M_\sun$ case in Table \ref{tab:MassRcgrid}. We chose an upper limit of 128 au for two reasons. One is that observations show that disc radii can be as high as 100-200 au \citep{2018ApJ...859...21A}. Observations also show that the centrifugal radius can be as high as 100-200 au even around young forming stars \citep{2014ApJ...791L..38S,2016ApJ...820L..34S}. Additionally, we find that if the centrifugal radius becomes very large ($\gg$ 100 au), the protoplanetary disc lifetime becomes much longer than the expected lifetimes of a few up to 10 Myr \citep{2001ApJ...553L.153H,2010A&A...510A..72F}, assuming $\alpha_\mathrm{min} = 10^{-2}$. 

After we have looked at the effect of the centrifugal radius on the disc evolution we also introduce a change in the cloud core mass and therefore also the resulting stellar mass. The other mass cases listed are in Table \ref{tab:MassRcgrid}. We use the 1 $M_\sun$ case as the reference and then scale the centrifugal radius range for the other cases linearly with the cloud core mass. This is based on the fact that the ratio between the rotational and gravitational energy of molecular cloud cores is independent of the cloud core size  \citep[e.g.][]{1993ApJ...406..528G, 2018ApJ...865...34C}. By setting this ratio to be the same for all cloud cores, one finds that the centrifugal radius of the protoplanetary disc scales linearly with the mass of a Bonnor-Ebert sphere. We note that from Eq. \eqref{eq:ac}  one might expect the centrifugal radius to scale as the inverse of the cloud core mass. However, the radius of a Bonnor-Ebert sphere scales linearly with its mass, giving a linear scaling between the centrifugal radius and cloud core mass (see Appendix \ref{app:BE_Rc_M} for a more detailed description).

\subsubsection{Monte Carlo cluster}

Once we have explored the effects of both the centrifugal radius and the mass of the cloud core we move on to using a Monte Carlo method for selecting the final centrifugal radius and the cloud core mass. With this we aim to mimic the evolution of a star-forming cluster for several million years. 
A total of 100 young stars and their protoplanetary discs are simulated. We do not model any interactions between the different star--disc systems, for example through gravity or gas dispersion due to radiation from nearby stars. That is, each star--disc system is modelled completely independently from the others, but we analyse them as a group.

We pick cloud core masses from a stellar IMF under the assumption that the entire mass of the cloud core will eventually end up in the star. For this we use the IMF of \cite{2001MNRAS.322..231K} in the mass range 0.1-2.0 $M_\sun$. The final centrifugal radius is picked from a normal distribution with a mean of 40 and a standard deviation of 35. This value is then scaled with the mass of the cloud core. We refer to Appendix \ref{app:BE_Rc_M} for the motivation behind this scaling.
  We further put a lower and upper limit on the centrifugal radius of 2 au and 120 au. Due to the scaling of centrifugal radius with cloud core mass this lower limit results in the low-mass cases being biased towards discs that are large and massive relative to the star. The lower limit is constrained by computational costs and the upper limit is set because discs with overly large centrifugal radii live much longer than the expected protoplanetary disc lifetimes as discussed in Sect. \ref{sec:res:grid:rc}. 
  
  Stellar clusters form from clumps inside giant molecular clouds \citep{2007ARA&A..45..565M}. Simulations suggest that the lifetimes of these clumps appear to range from  $10^5$ to $10^7$ yr \citep{2011ApJ...735...99F}. To mimic the stars forming at different times we give a time delay of between 0 and 1 Myr for the collapse of each cloud core from when the cluster begins to form.

Other distributions of the centrifugal radius within the same lower and upper limit were tested, including a normal distribution with different mean and standard deviation, and a uniform distribution,  but no significant differences were found on the cluster evolution. We do find that a distribution that is weighted towards higher centrifugal radius cases gives a larger average dust disc mass. However, the age spread between the discs also significantly affects the average dust mass, and this is a parameter that can have a large spread.

\subsection{Method}

We evolve the disc using a one-dimensional numerical model similar to that of \cite{1981MNRAS.194..967B} and \cite{2017MNRAS.469.3994B}. The computational domain is divided into 100 logarithmically spaced cells.  The outer edge is 10 000 au and the inner edge is set such that it is 1/100 of the final centrifugal radius. Gas and dust that lands inside the inner edge of the domain is treated as being directly accreted onto the star. 

%

\section{Results}
\label{sec:results}

\subsection{Reference case}
\label{sec:results:ref}

\begin{figure}[t]
   \resizebox{\hsize}{!}{\includegraphics[width=\hsize]{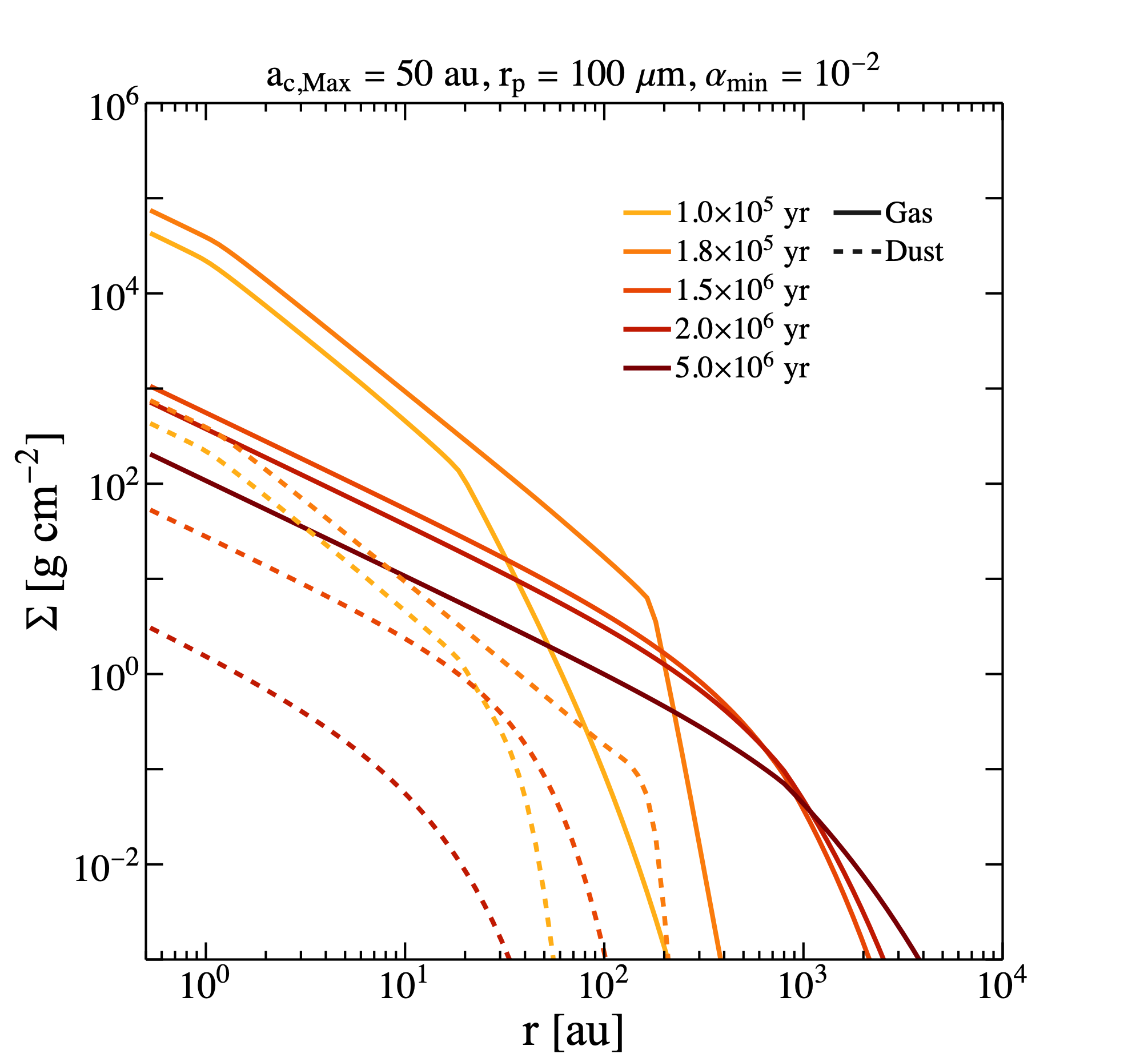} }
      \caption{ Snapshots of gas (solid lines) and dust (dashed lines) surface densities of the  reference case.  In the evolution phase, the inner disc is drained of gas as it is accreted by the star and the outer disc expands as gas is transported outwards to conserve angular momentum. In the first $\approx$ 1.0 Myr the dust surface density mimics that of the gas in most of the disc out to radii of about 100 au. Beyond this radius, radial drift of dust is already rapid, draining the dust disc. When the disc is  between 2 and 2.5 Myr old, the dust disc is completely drained.
      }
         \label{Fig:Nom:Surfdens}
   \end{figure}

First we investigate the reference case with parameters as explained in Sect. \ref{sec:sims}.
We numerically evolve the gas and dust surface density. Therefore, we begin by examining the evolution of these two quantities. Figure \ref{Fig:Nom:Surfdens} shows snapshots in time of the gas (solid lines) and dust (dashed lines).

We define the formation phase of the disc as the time period during which there is mass infalling from the collapsing cloud core onto the disc, that is, the time it takes for the cloud core to collapse. The collapse of a 1 $M_\sun$ cloud core at a temperature of 10 K lasts about $1.8\times 10^5$ yr. This depends on the temperature of the cloud core, as it affects the size of the Bonnor-Ebert sphere. 
 
 The gas surface density in the disc during this formation phase changes both by the infalling material and due to viscous evolution, but the former dominates the mass evolution for most of this phase. 
  Initially, this leads to high surface densities in the inner parts of the disc, almost reaching $10^5\ \mathrm{g} \ \mathrm{cm}^{-2}$ at the inner edge of our simulation domain (0.5 au). Towards the end of the formation phase the disc becomes gravitationally unstable at radii from $r\approx\! 2$ au to $r\approx\! 200$ au. In these gravitationally unstable regions the viscosity increases because of gravitational torques according to the prescription in Eq. \eqref{eq:alpha} and \eqref{eq:alpha_q}. The increased viscosity leads to more efficient transport of gas compared to the stable regions. 
 
 Once the formation phase is over, the gas disc only evolves viscously.  
 With constant particle size and density, the Stokes number is a function of the gas surface density only (assuming the Epstein drag regime, see Eq. \ref{eq:Epstein}). Therefore, as long the gas surface density is sufficiently high, the Stokes number is kept low and dust drift becomes negligible. In this reference case it takes about 0.75 Myr for the gas surface density to decrease enough in the outer disc for the Stokes number to increase to the point where radial drift becomes significant. From this point, the dust disc is drained in about 1.75 Myr. This can be seen in the 2.5 Myr snapshot of Fig. \ref{Fig:Nom:Surfdens}.

\begin{figure}[t]
\resizebox{\hsize}{!}{\includegraphics[width=\hsize]{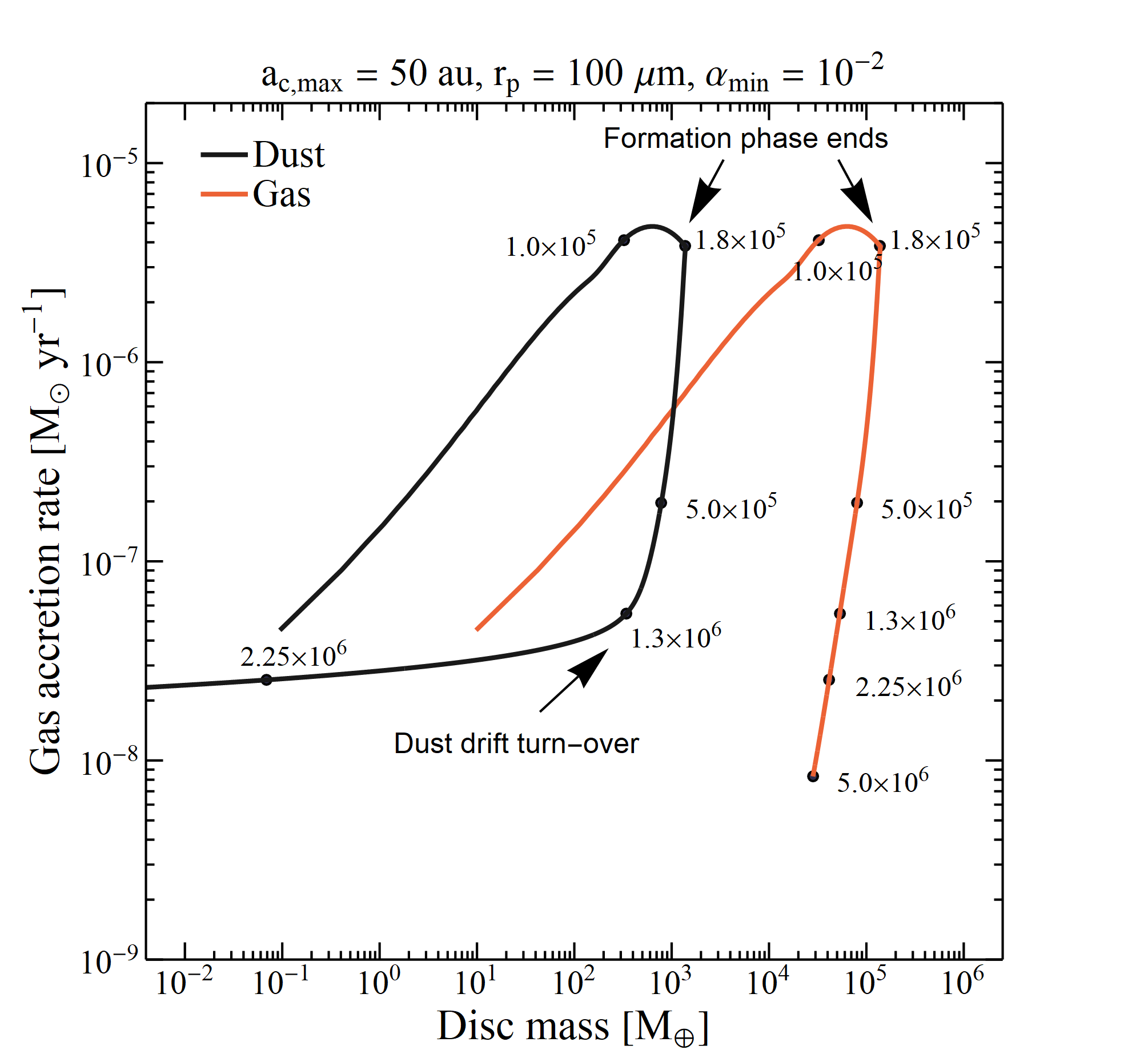} }
      \caption{  Stellar gas accretion rate as a function of disc mass for the  reference case. The black curve shows the dust disc mass and the red curve shows the gas disc mass. The black dots mark the age in years as labelled. There are three clear regimes in this relation. During the formation phase of the disc, the gas accretion rate and the disc dust mass are both increasing. Once the formation phase has ended, dust drift is slow and the gas accretion rate drops faster than the disc dust mass. Rapid draining of the dust disc mass begins around  $t \approx\! 1.3 \times 10^6$ and lasts about 1 Myr. 
      }
         \label{Fig:Nom:MdotvsMdisc}
   \end{figure}

We now explore how the stellar gas accretion rate evolves with the mass in the disc. This is shown in Fig. \ref{Fig:Nom:MdotvsMdisc} for the reference case.  There are three clear regimes for the dust disc and two for the gas disc, starting with the disc formation phase. During the first 0.1 Myr of this regime, both the stellar gas accretion rate and the dust-disc mass are increasing. In the remainder of the formation phase the gas accretion does not change significantly. It is during this later part of the formation phase that gravitational instability  is active. The gas accretion rate peaks at $\dot{M}_\mathrm{g}\approx\! 5\times 10^{-6}\ M_\sun \ \mathrm{yr}^{-1}$. The dust mass reaches a maximum mass of $M_\mathrm{d}\approx\! 1.4\times 10^3 \ M_\Earth$ and the gas mass reaches  $M_\mathrm{g} \approx\! 0.4 $ $M_\sun$. 
 This far in the disc evolution, the dust-to-gas ratio remains close to $\Sigma_\mathrm{d}/\Sigma_\mathrm{g} \approx\! 0.01$ throughout the disc. 
 
The second regime begins once the formation phase is over. During this regime the gas surface density is still high and as a consequence the Stokes numbers are low. Dust drift is therefore slow, leading to a very slow depletion of the dust disc. 
This regime last until about 1.3 Myr. 
 
 Subsequently, the dust clearing regime begins, which can be identified by the dust drift turn-over in Fig. 2. Now, the lower gas surface densities result in increased Stokes numbers of the particles and radial drift of dust can rapidly reduce the mass of the dust disc. At an orbital radius of 1 au, the Stokes number increases to $\tau_\mathrm{s} \approx\! 10^{-4}$ at $t = 2$ Myr. 
After 2.25 Myr the dust disc has been drained to less than 0.1 $M_\Earth$.

In summary, we find that a single disc goes through three regimes in gas accretion rate versus dust disc mass. The first is the formation phase where both the accretion rate and the dust disc mass grow. After the formation phase is over, a phase of slow radial dust drift is entered.   Eventually, the gas surface density drops and the Stokes number becomes large. Radial dust drift now becomes rapid and the dust disc is drained.

\subsection{Dependency on the centrifugal radius}
\label{sec:res:grid:rc}
\begin{figure}[t]
\resizebox{\hsize}{!}{
\includegraphics[width=\hsize]{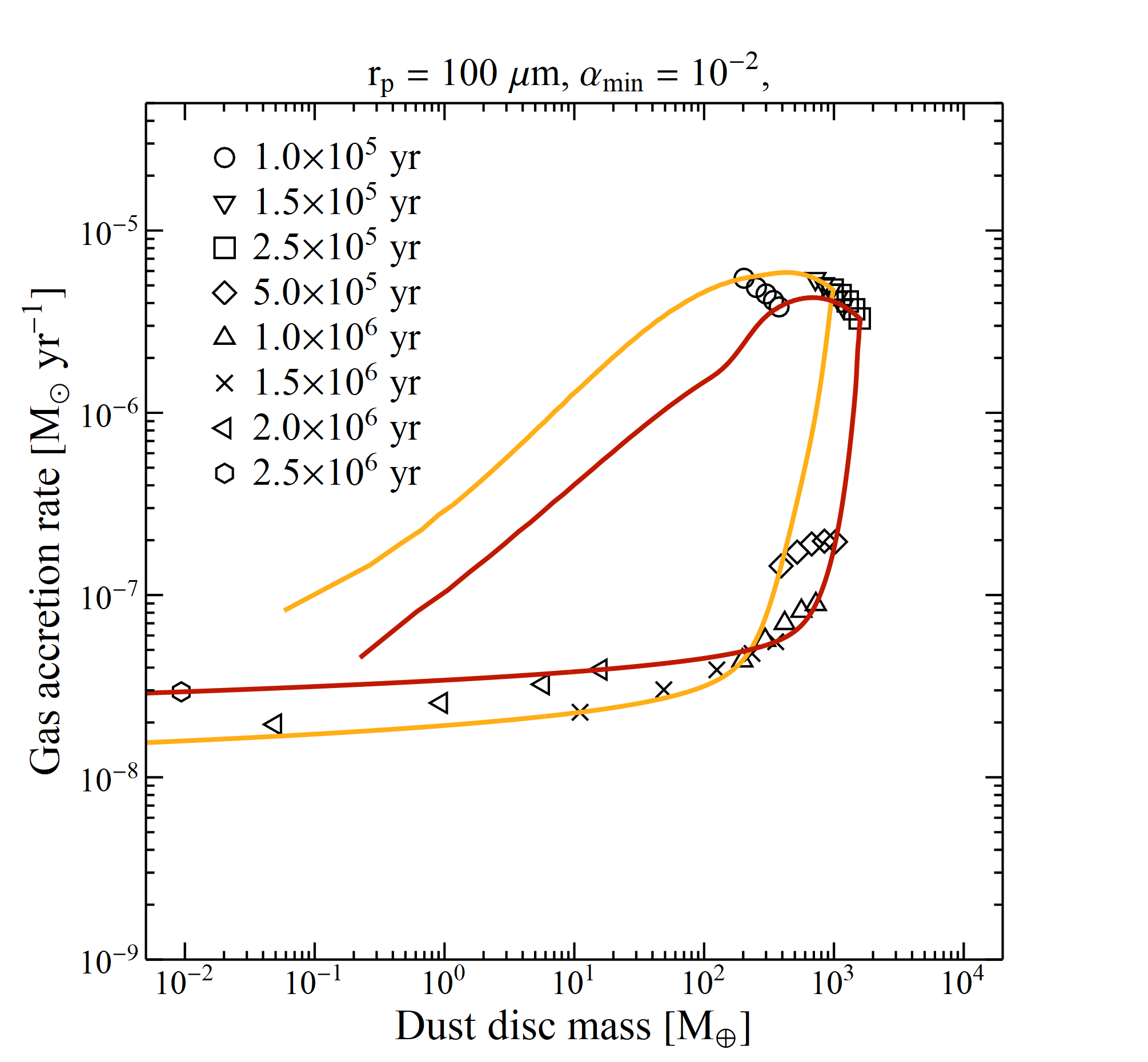}} 
\caption{  Stellar gas accretion rate and dust disc mass as a function of time, for discs with final centrifugal radii of  8, 16, 32, 64, and 128 au forming from 1 $M_\sun$ cloud cores. The red and yellow lines show the tracks of the two cases with the highest and lowest final centrifugal radius, respectively. The formation phase ends at $t\approx \!1.8\times 10^5$ yr. During the formation phase a trend can be seen where systems with smaller centrifugal radii have lower dust disc masses and slightly higher gas accretion rates. 
Between $1.8\times 10^5$ yr and $5.0\times 10^5$ yr the gas accretion rates drop significantly, while dust disc masses decrease only slightly. Discs with a smaller final centrifugal radius experience a larger drop in accretion rate.  A lower final centrifugal radius leads to faster evolution of the protoplanetary disc. At $2.5\times 10^6$ yr, only the case with the highest final centrifugal radius remains, but the dust disc mass is only $\approx \! 10^{-2}$ $M_\Earth$.}
         \label{Fig:Grid:MdotvsMdisc1Ms}
\end{figure}

In this section, we explore the effect of the centrifugal radius on the evolution of the protoplanetary disc. For a given mass of the cloud core, a change in the rotation frequency of the core, and thus the centrifugal radius, affect the evolutionary track of the disc in the  phase space of stellar gas accretion rate versus dust disc mass.
This is shown in Fig. \ref{Fig:Grid:MdotvsMdisc1Ms}. The red and yellow lines show the evolutionary track of the highest and lowest centrifugal radius case, respectively. The different symbols represent different times as shown in the legend. For each symbol, the five points show the five final centrifugal radii cases of the 1 $M_\sun$ set in Table \ref{tab:MassRcgrid}. 

 During the formation phase, the case with the smallest centrifugal radius (yellow line) has the highest gas accretion rate and the lowest dust mass of all cases.  This trend holds until the end of the formation phase ($t \approx 1.8\times 10^5 \ \mathrm{yr}$).
  As the discs evolve past the formation phase, a disc with a lower final centrifugal radius evolves faster. Therefore, it will reach the dust drift turn-over earlier and will drain its dust faster.  

\subsection{Dependency on cloud mass}
\label{sec:res:grid:mass}

\begin{figure*}[t]
\centering
\includegraphics[width=17cm]{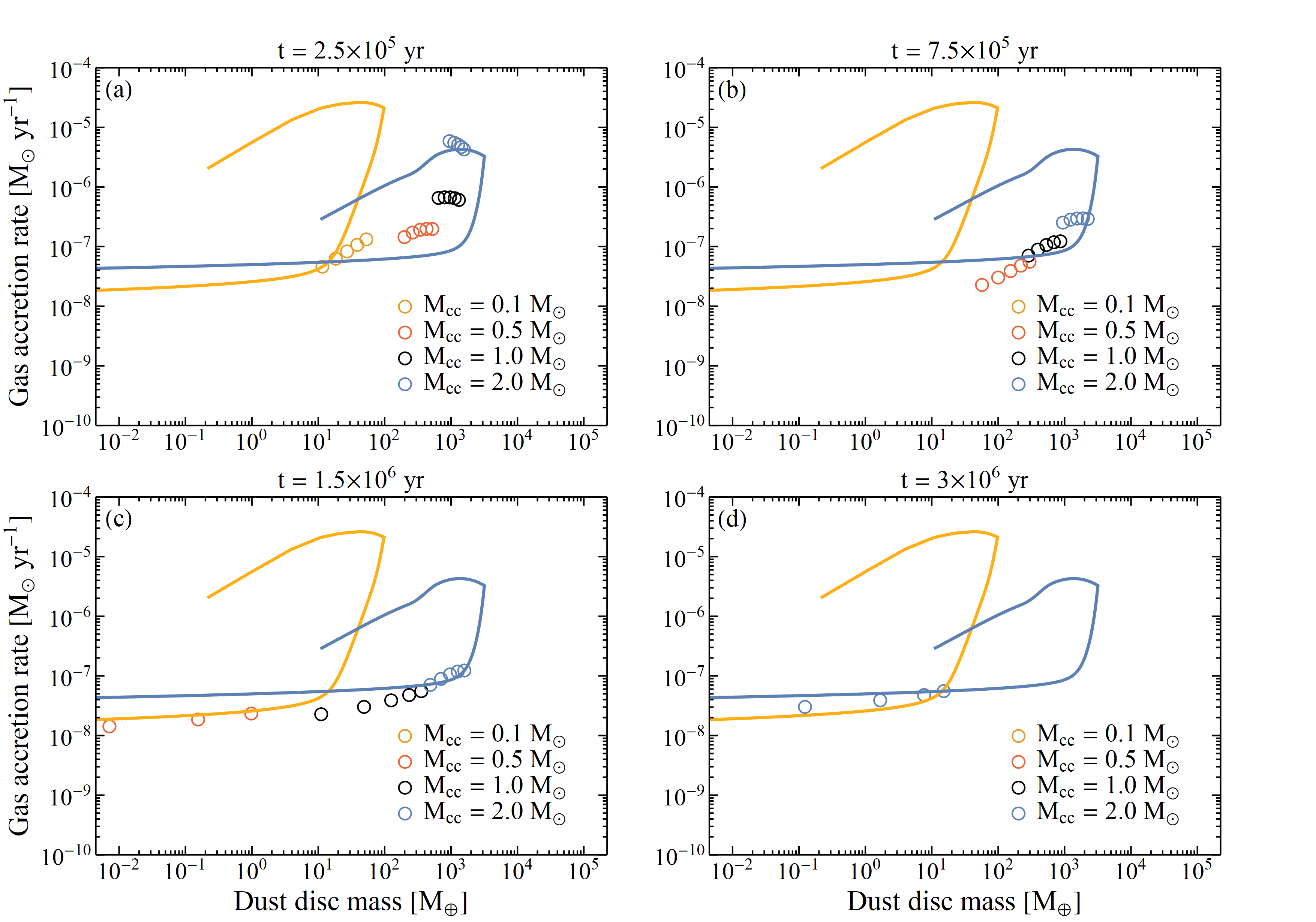} 
\caption{  Gas accretion rate as a function of dust disc mass for protoplanetary discs formed from cloud cores with masses and final centrifugal radii shown in Table \ref{tab:MassRcgrid}. The yellow and blue lines trace the time evolution of the cases with the lowest and highest mass, and lowest and highest centrifugal radius, respectively.  As a general trend, discs from low-mass cloud cores evolve faster than those from higher-mass cloud cores. However, the speed with which the disc evolves also depends on the final centrifugal radius.
At 3 Myr (panel (d)), all dust discs except for the ones coming from 2 $M_\sun$ cloud cores have been completely drained out.}
         \label{Fig:Grid:MdotvsMdiscMassRcGrid}
\end{figure*}

We now investigate the effect of changing the mass of the cloud core, and therefore also the final stellar mass. We use the cases shown in Table \ref{tab:MassRcgrid}.  For a given cloud core temperature, the timescale for the formation phase of the disc increases nearly linearly with the mass of the cloud core.

Again, we look at the stellar gas accretion rate as a function of dust disc mass. This is shown in Fig. \ref{Fig:Grid:MdotvsMdiscMassRcGrid}. Panels (a), (b), (c), and (d) show the discs at 0.25, 0.75, 1.5, and 3.0 Myr. Unsurprisingly, the more massive the cloud is, the more massive the disc becomes. However, this is not a strict relation, given our spread in centrifugal radii. Given a sufficiently high centrifugal radius from a lower-mass case, and a sufficiently low centrifugal radius from a higher-mass case, it is possible to find cases where the higher-mass core ends up with a lower-mass disc than the low-mass core. The reason for this is that with a high centrifugal radius, more mass is deposited onto the disc rather than directly onto the star.
This can be seen in panels (a) and (b) of Fig. \ref{Fig:Grid:MdotvsMdiscMassRcGrid}, where the 0.5 $M_\sun$ and 1.0 $M_\sun$ cases overlap in disc dust mass. 

When the system is 0.75 Myr old (panel (b) of Fig. \ref{Fig:Grid:MdotvsMdiscMassRcGrid}), we can see that the dust discs from the 0.1 $M_\odot$ core have been depleted of dust and that all the remaining discs line up, showing a trend of higher cloud core masses giving higher gas accretion rates and higher disc masses.  As a general trend, the lower the beginning cloud core mass, the earlier the dust drift turn-over is reached. Because the discs from low-mass cloud cores reach the point of rapid radial drift earlier, they are drained of dust faster. Therefore, at the 1.5 Myr mark (panel (c) of Fig. \ref{Fig:Grid:MdotvsMdiscMassRcGrid}), the discs from the lower-mass cloud cores have much less massive dust discs. At 3 Myr (panel (d)) only the discs  from the 2 $M_\sun$ still have a dust component remaining. The most massive of these two discs had a total mass of $M_  \mathrm{d} \approx\! 3\times 10^3$ $M_\Earth$ at the end of the formation phase. This disc is drained to less than $10^{-1}\ M_\Earth$ in 3.5 Myr.

To briefly summarise the present section and the previous one, we find that both gas and dust discs evolve faster for lower mass cloud cores (therefore those with lower resulting stellar mass) with smaller final centrifugal radii
. As a result, the dust disc is drained faster around lower mass stars and in discs with lower final centrifugal radii.

\subsection{Monte Carlo cluster}

In this section, we continue to explore the relationship between the  stellar gas accretion rate to disc mass, but now we use a Monte Carlo method to select the cloud core mass and the final centrifugal radius, with the aim of mimicking the evolution of a star-forming region for several million years. The parameter space we study is specified in Sect. \ref{sec:sims}.

\begin{figure*}[t]
\centering
\includegraphics[width=17cm]{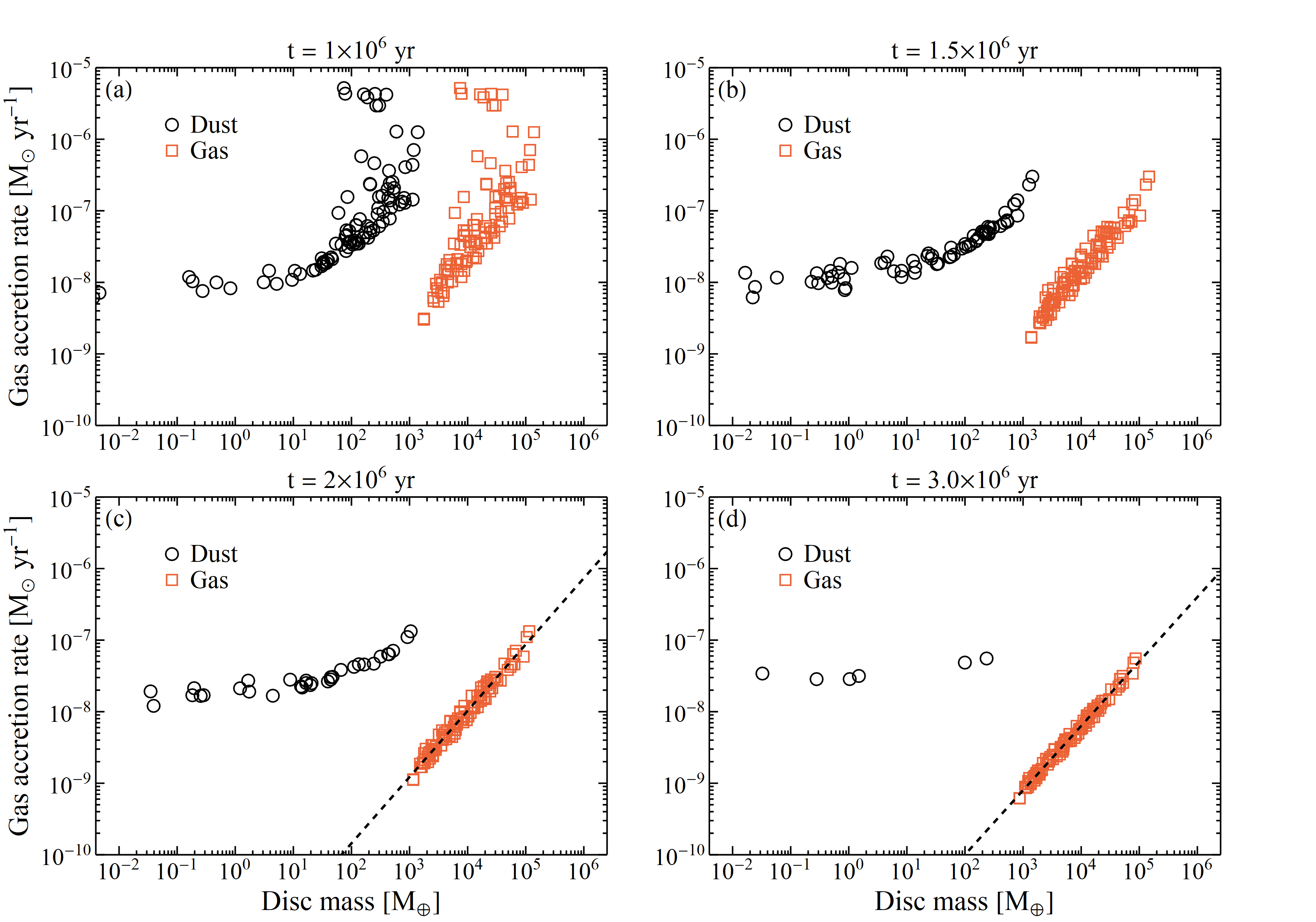} 
\caption{Gas accretion rate as a function of dust (black circles) and gas (red squares) disc mass for protoplanetary discs of the stellar cluster. The dashed black lines show power-law fits of the accretion rate as a function of the gas disc mass ($\dot{M}\propto M_\mathrm{g}^{0.9}$). The discs are given a random individual age of between 0.0 and 1.0 Myr to mimic a spread in formation times. The panels going from (a) to (d) show the protoplanetary discs at the cluster ages of 1.0, 1.5, 2.0, and 3.0 Myr. For the first 1 Myr the dust is coupled well to the gas in most discs, and dust disc masses evolve in a very similar way to the gas disc mass. Some of the lowest mass discs have reached the dust drift turn-over at this point. Beyond this time more and more of the dust discs reach the dust turn-over and begin to drain rapidly. At 1.5 Myr the lower mass discs have reached the turn-over point, as can be seen in the flattening of the low-mass end  of the dust discs in panel (b). At 2 Myr only the most massive dust discs are yet to reach the turn-over and at 3 Myr nearly all discs are drained of dust. Only 4 out of 100 discs still have a dust mass greater than 1 $M_\Earth$, although two of these have a mass of more than 100 Earth masses. 
      }
         \label{Fig:MC:MdotvsMdisc}
\end{figure*}

Figure \ref{Fig:MC:MdotvsMdisc} shows the stellar gas accretion rate versus the disc gas and dust mass for the Monte Carlo cluster. The dust is shown in black circles and the gas in red squares. Panels (a) to (d) show the cluster at 1, 1.5, 2 and 3 Myr respectively. At 1 Myr, the majority of the discs have left the formation phase and are in the phase of slow dust drift. A few remain in the formation phase and are characterised by low disc masses  and/or high accretion rates, and some of the lowest mass discs have passed the dust drift turn-over. At this point, the dust traces the gas very closely, at one-hundredth of the gas mass, since it is still tightly coupled to the gas in all but the least massive discs. At 1.5 Myr, all discs have left the formation phase (for the range of parameters used here this is inevitable: the most massive cores allowed take $\approx \! 0.4$ Myr to collapse and the age spread between cores is only 1 Myr). Most discs are now near or at the dust drift turn-over. The lowest mass discs with the lowest centrifugal radii have passed the dust drift turn-over and the low-mass end of the dust track is now beginning to flatten out.
 When the cluster is 2 Myr old, nearly all discs have passed the dust drift turn-over and are in the regime of rapid radial dust drift, and many of the low-mass discs are devoid of dust. At 3 Myr we see that the majority of dust discs are drained, and of those that remain, only 4 out of 100 have a total mass of more than 1 $M_\Earth$.
 
Once all the discs have left the formation phase, the gas component settles into a fairly narrow band in the accretion-rate--disc-mass space. This can be seen in panels (b), (c), and (d) of Fig. \ref{Fig:MC:MdotvsMdisc}. As the discs evolve forward in time, both the accretion rate and the disc masses decrease as the gas is accreted onto the star. The spread in the relation also decreases with time. That gas discs end up in such a tight space in the gas-accretion-rate-versus-disc-mass relation was also found theoretically by \cite{2017MNRAS.472.4700L}. 

\section{Scaling relations}
\label{sec:scaling}

In this section we explore various scaling relations between the gas accretion rates, the disc, and the stellar masses for the reference case and the cluster of discs. We then compare the scaling relations in the cluster to observations. For the single disc and the cluster, the scaling relations are summarised in Tables \ref{tab:scaling_relations_single} and \ref{tab:scaling_relations_cluster}, respectively.

\subsection{The scaling of the mass accretion rate with the disc mass}
Looking at the single star in the reference case, we find that during the formation phase the gas accretion rate scales with both gas and dust disc mass as $\dot{M} \propto M_\mathrm{d,g}^{0.5}$. After the formation phase but before the dust drift turn-over we find $\dot{M} \propto M_\mathrm{d}^{4.0}$. After the turn-over we find  $\dot{M} \propto M_\mathrm{d}^{0.04}$. For the gas disc we find  $\dot{M} \propto M_\mathrm{g}^{3.0}$ in the post turn-over phase. 

Now we examine the cluster of discs. In the gas accretion rate versus dust disc mass we do not find that a power law describes the data well at 2 Myr. At 3 Myr most of the discs are drained of dust, so we chose not to make a fit to the few remaining discs. A broken power law would likely provide a better fit than a single power law; one would represent discs that have not yet reached the dust drift turn-over, and the other
would represent those that have passed it. 

Observations report a near-linear scaling between the stellar accretion rate and the dust disc mass, $\dot{M} \propto M_\mathrm{d}^{0.7}$ \citep{2016A&A...591L...3M}, most similar to early times in our Monte Carlo run ($\lesssim 1.5$ Myr). These observations nevertheless also show a spread of several orders of magnitude in the gas accretion rate for a given dust mass. For an evolved cluster of discs, we find a much tighter relation with a spread that is only a factor of a few at most. The larger spread in observed discs could be due to a combination of different disc ages, intrinsic viscosity variations, and an intrinsic spread in dust sizes. We explore the effects of the viscosity and the dust size in Appendices \ref{app:alpha} and \ref{app:dustsize}. Another difference is that we do not reach  accretion rates as low as $<10^{-9}$ $M_\sun$ yr$^{-1}$; such low accretion rates are common in the sample of \cite{2016A&A...591L...3M}. However, our model does not include gas mass-loss mechanisms such as disc winds, disc photoevaporation, or accretion of gas onto planets. These could have the effect of decreasing the gas accretion rate for a given dust disc mass.

In the gas-accretion-rate-versus-gas-disc-mass relation in the cluster we find a near-linear scaling with $\dot{M} \propto M_\mathrm{g}^{0.9}$ at 2 Myr and 3 Myr. This can be seen in the black dotted lines in panels (c) and (b) of Fig. \ref{Fig:MC:MdotvsMdisc}. A common assumption made in observational studies is that the gas mass is 100 times the dust mass. However, at 2 and 3 Myr most or all of the discs in our model are in the phase of rapid radial dust drift where this assumption does not hold. Hence, a scaling relation measured for the dust mass does not necessarily translate directly onto the gas mass. 
Directly comparing the gas-accretion-rate-to-gas-disc-mass relation to observations is complicated because of the uncertainty in estimating the mass locked in H and He gas based on CO spectral line emission \citep{2016A&A...585A.136M}.

\begin{figure}[t]
\centering
\includegraphics[width=\hsize]{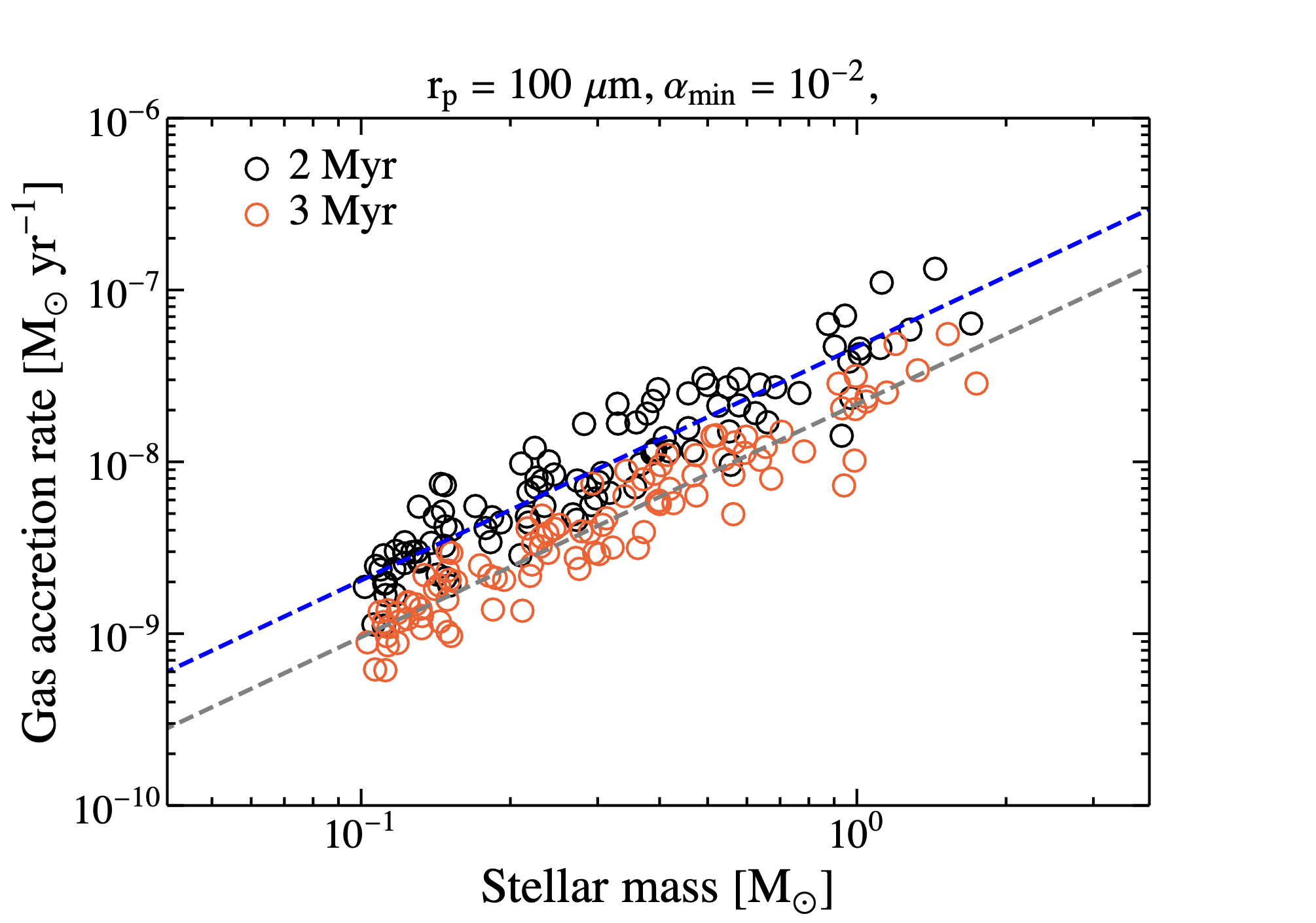} 
\caption{Gas accretion rate as a function of stellar mass of the cluster of discs at 2 Myr (black) and 3 Myr (red). The blue and grey dashed lines are fits to the data at 2 Myr and 3 Myr respectively. Both scaling relations have a slope of $\approx\! 1.4$.
      }
         \label{Fig:MC:MdotvsMstar}
\end{figure}

\subsection{The scaling of the gas mass and stellar accretion rate with stellar mass.}

  The gas disc mass scales slightly steeper than linear with the stellar mass in the cluster. We find $M_\mathrm{g} \propto M_\star^{1.4}$ at 2 Myr and 3 Myr. The gas accretion rate scales nearly linearly with stellar mass. At 2 Myr and 3 Myr we find $\dot{M} \propto M_\star^{1.4}$. This is shown in Fig. \ref{Fig:MC:MdotvsMstar}. With a single power-law fit, a slope of 2.2 has been found. However, this relation appears to show a broken power-law scaling with a much shallower slope of 0.56 for $M_\star > 0.3\ M_\sun$  (and a much steeper slope of 4.3 below 0.3 $M_\sun$)  \citep[]{2017A&A...604A.127M, 2017A&A...600A..20A}.
   A direct comparison is therefore difficult at this point. Nonetheless, Fig. \ref{Fig:MC:MdotvsMstar}  shows no broken scaling. However, at the high-mass end, the slope that we find (1.4) sits between the single power-law slope (2.2) and the high-mass end of the broken power-law slope (0.56) of \cite{2017A&A...604A.127M}.

\begin{figure}[t]
\centering
\includegraphics[width=\hsize]{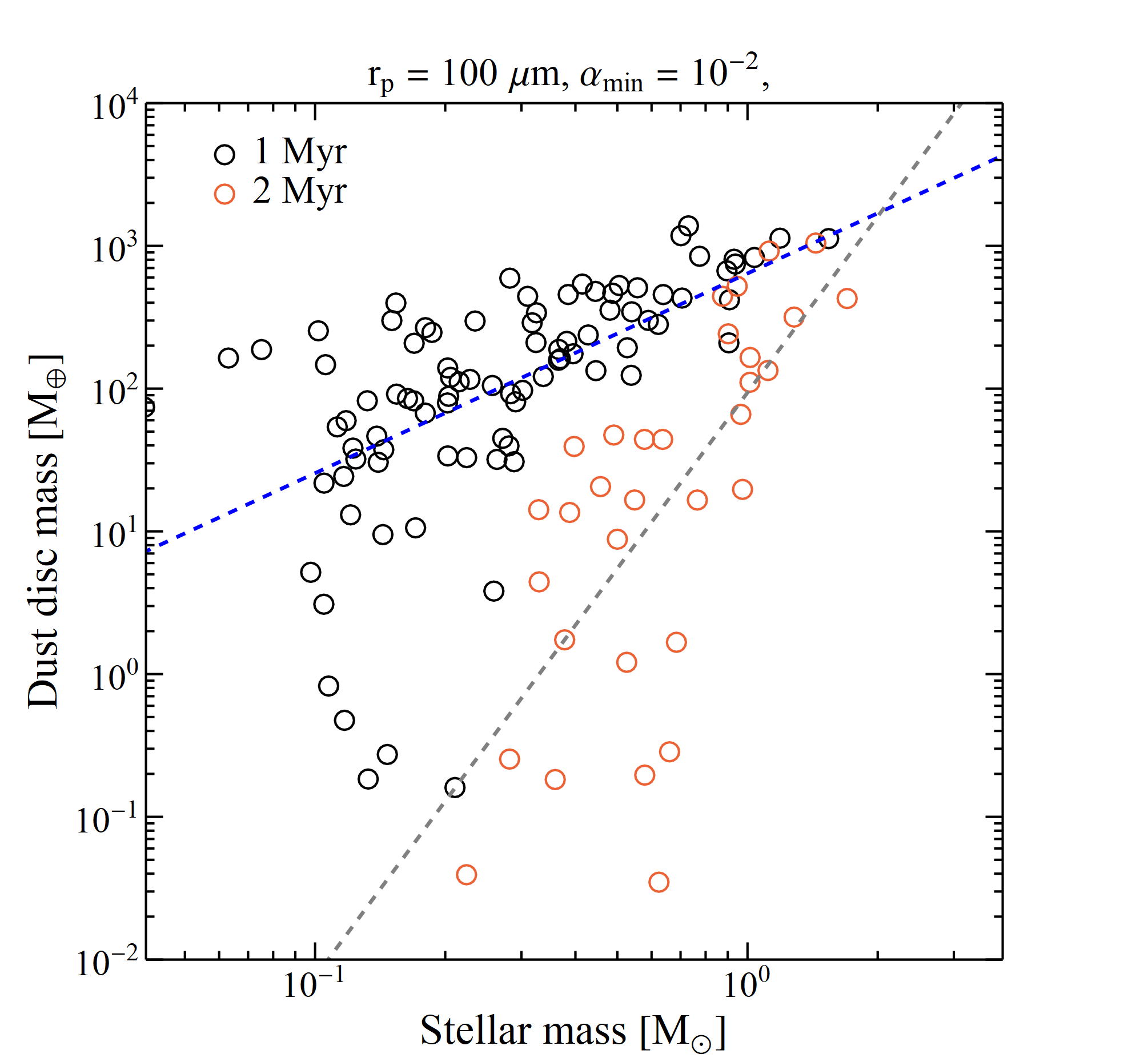} 
\caption{Dust disc mass as a function of the stellar mass at 1 Myr (black) and 2 Myr (red). The dashed blue and grey lines show fits to the data. The scaling relation steepens with time from a slope of 1.4 at 1 Myr and 4.1 at 2 Myr.
      }
         \label{Fig:MC:MdustvsMstar}
\end{figure}

\subsection{The scaling of the dust mass with the stellar mass}
 
 In terms of how the dust mass scales with the stellar mass, the situation is slightly more complicated.  The dust disc mass as a function of stellar mass is shown in Fig. \ref{Fig:MC:MdustvsMstar}. At 2 Myr, we  find $ M_\mathrm{d} \propto M_\star^{4.1}$ if all discs with a dust mass of less than $10^{-2}$ $M_\Earth$ are excluded from the fit. This is the criterion we have previously used to define when a disc is drained. If instead we exclude discs less massive than $10^{-1}$ $M_\Earth$ , we find $M_\mathrm{d} \propto M_\star^{3.8}$. Hence, the slope does not appear sensitive to the cut off we use. 
 
If alternatively we look at when the cluster is 1.5 Myr old and apply the same criteria, we find 
 $M_ \mathrm{d} \propto M_\star^{3.1}$ when excluding discs below  $10^{-2}$ $M_\Earth$. When the cluster is 1 Myr old we find $M_\mathrm{d} \propto M_\star^{1.4}$.
 
Comparing this to observations, the slope we find is steeper than expected. \cite{2016ApJ...831..125P} found $M_\mathrm{d} \propto M_\star^{1.1-1.6}$ in the 1-2 Myr-old Taurus star-forming region, $M_\mathrm{d} \propto M_\star^{1.1-1.8}$ in the 1-3 Myr-old Lupus region, $M_\mathrm{d} \propto M_\star^{1.3-1.9}$ in the 2-3 Myr-old  Chamaeleon I,  and finally $M_\mathrm{d} \propto M_\star^{1.9-2.7}$  in the 10 Myr-old Upper Sco region. Similarly, \cite{2016ApJ...827..142B} found a scaling in Upper Sco of  $M_\mathrm{d} \propto M_\star^{1.67\pm0.37}$. \cite{2016ApJ...831..125P} suggest that the difference in slope between the first three regions is due to the lower sensitivity of  the survey of Taurus, and that the scaling is similar for star-forming regions between 1 and 3 Myr. Only when our cluster is 1 Myr old does the scaling we find agree with observations.

Our findings and those of \cite{2016ApJ...831..125P} both suggest that the scaling between stellar mass and dust disc mass becomes steeper as the cluster becomes older. Although, in our model the steepening appears  to occur significantly faster. \cite{2016ApJ...831..125P} suggested that the steepening of the relation can be explained when dust growth is fragmentation limited, as this leads to more rapid dust clearing around low-mass stars. Similarly, in our model we find that discs from lower-mass cloud cores, and consequently around lower-mass stars, are more quickly drained of dust. This leads to a steepening of the scaling relation. Our findings of lower-mass stars having dust discs that live shorter do however seem to contradict observations that report a larger fraction of discs around lower mass stars than around higher mass stars in Upper Sco and the 5-12 Myr old Collinder 69 cluster \citep{2006ApJ...651L..49C,2012A&A...547A..80B}.

\begin{table}[]
\centering
\caption{Scaling relations between various disc and stellar parameters for a single disc evolving over time. The values indicate the slope, $a$, of the power law.}
\label{tab:scaling_relations_single}
\begin{tabular}{llll}
\hline
                            & Formation phase& Pre turn-over & Post turn-over \\ \hline
$\dot{M}\propto M_\mathrm{d}^a $ & 0.5             & 4.0                      & 0.04                     \\
$\dot{M}\propto M_\mathrm{g}^a $ & 0.5             & 3.0                      & 3.0                     \\ \hline
\end{tabular}
\end{table}

\begin{table}[]
\centering
\caption{Scaling relations between various disc and stellar parameters for a cluster of discs at 2 and 3 Myr. The values indicate the slope, $a$, of the power law. 
}
\label{tab:scaling_relations_cluster}
\begin{tabular}{llll}
\hline
                &    1 Myr    & 2 Myr & 3 Myr \\ \hline
$\dot{M}\propto M_\mathrm{g}^a$ & & 0.9  & 0.9   \\
$M_\mathrm{d}\propto M_\star^a$ & 1.4 & 4.1  &    \\
$M_\mathrm{g}\propto M_\star^a$ & & 1.4  & 1.4   \\
$\dot{M}\propto M_\star^a$  &  & 1.4  & 1.4   \\ \hline
\end{tabular}
\end{table}

%

\section{Discussion}
\label{sec:discussion}

\subsection{Observed disc masses}

In the Monte Carlo parameter exploration we find dust disc masses up to $\approx\! 2.3\times 10^3$ $M_\Earth$ and gas disc masses up to $\approx\!  0.7$ $M_\sun$. The most massive disc comes from the most massive cloud core with the highest angular momentum budget at the end of its formation phase, when the disc is 0.34 Myr old. At a cluster age of 1 Myr we find dust masses ranging from zero, i.e. a completely drained disc, to $\approx\! 1.3\times 10^3$ $M_\Earth$, and gas masses between $\approx\! 3\times 10^{-3}$ $M_\sun$ and $\approx\! 0.4$ $M_\sun$. Once the cluster is 2 Myr old, many dust discs have started to drain and several of the lowest mass discs are completely drained. At this age we find dust masses up to $\approx\! 1\times 10^3\ \mathrm{M}_\Earth$ and gas masses are in the range $\approx\! 4 \times 10^{-3}$ $M_\sun$ to $\approx\! 0.3$ $M_\sun$.  

 Observational measurements from CO and HD put gas disc masses in the range of $\approx\! 10^{-5} \ M_\sun - 10^{-1}$ $M_\sun$ \citep{2013Natur.493..644B,2016ApJ...828...46A,2016ApJ...831..167M,2017A&A...599A.113M}. However, it should be noted that gas masses measured from CO are very uncertain; see e.g. Fig. 3 in \cite{2016A&A...591L...3M}. This puts the lower limit of our gas masses a few orders of magnitude above the lowest observed values and our highest value a factor of a few higher than the observed sample.

\begin{figure}[t]
   \resizebox{\hsize}{!}{\includegraphics[width=\hsize]{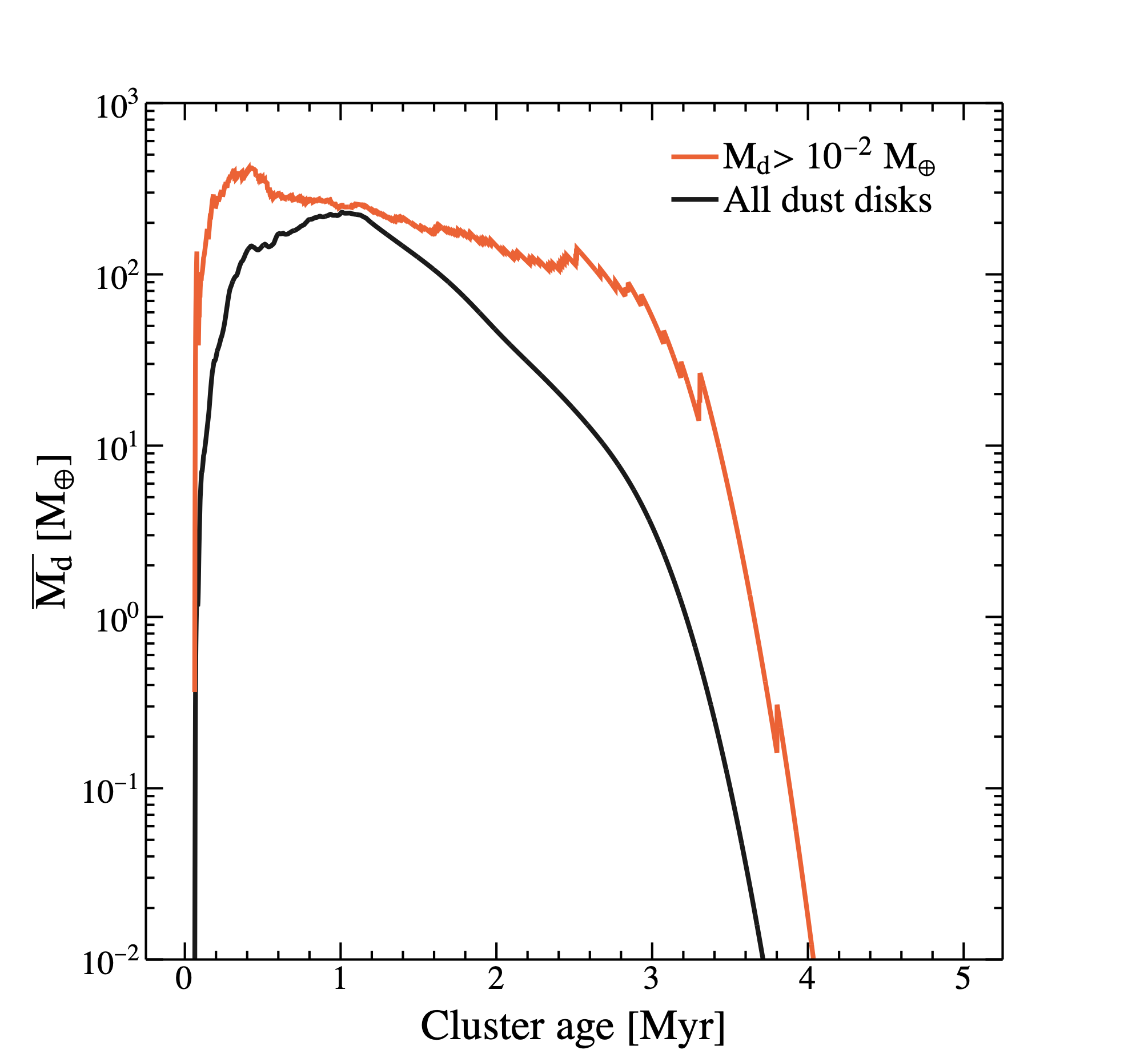} }
      \caption{Mean dust disc mass as a function of cluster age. The black curve shows the mean mass when including all dust discs and the red curve shows the mean mass when dust discs with masses $<10^{-2}$ $M_\Earth$ are excluded.}
         \label{Fig:MdMean}
   \end{figure}

We can now identify at what ages the mean dust mass in our cluster of discs matches the mean dust mass in these observed star-forming regions. In both the 1-3 Myr-old Lupus  and the 1-2 Myr-old Taurus star-forming regions, dust disc masses range between $\approx\! 0.1$ $M_\Earth$ and $\approx\! 100$ $M_\Earth$, and the average mass is $\approx\! 15$ $M_\Earth$ \citep{2016ApJ...828...46A}.  A smaller average of 5 $M_\Earth$ was found in the 10 Myr-old Upper Sco star-forming region.
The average dust disc mass as a function of time  in our population synthesis model is shown in Fig. \ref{Fig:MdMean}. When including all dust discs regardless of their mass we find that our cluster has an average dust mass of 15 $M_\Earth$ at an age of 2.5 Myr, which is the average in Lupus and Taurus. This agrees well with the estimated ages of 1-3 Myr for Lupus and 1-2 Myr for Taurus. We find an average dust mass of 5 $M_\Earth$ at 2.9 Myr. If we exclude discs with masses $< 10^{-2}$ $M_\Earth$ we find an average dust disc mass of 15 $M_\Earth$ at 3.3 Myr, which again agrees well with observations. Figure \ref{Fig:MdMean} also shows that the mean dust mass declines from $\approx\! 200$ $M_\Earth$ at 1 Myr to $\approx\! 3$ $M_\Earth$ at $\approx\! 3$ Myr. This relatively rapid decline does not agree well with the retention of dust in the 10 Myr-old Upper Sco region, which has a mean mass of 5 $M_\Earth$. However, as with the scaling relations, this result is sensitive to the age spread given to the disc. An age spread of 2 Myr instead of 1 Myr gives a mean dust mass of 15 $M_\Earth$ at 3.5 Myr when all discs are included, 1 Myr later than when the age spread is 1 Myr.

There are some indications that protoplanetary dust disc masses may be underestimated. For example, \cite{2018ApJ...868...39G} investigated the effect of self-obscuration in protoplanetary discs due to them being partially optically thick. These latter authors find that protoplanetary disc masses may be underestimated by a factor of ten or more. Recent work by \cite{2019AJ....157..144B} also showed that protoplanetary disc masses may be underestimated. These authors used radiative transfer models to fit the SED of Taurus discs and found that masses may be one to five times larger than previously thought. Another recent study has shown that dust scattering is capable of reducing the emission from an optically thick disc, making it appear optically thin, and therefore leading to an underestimation of its mass \citep{2019ApJ...877L..18Z}. Furthermore, the previously mentioned discrepancy between the masses of protoplanetary discs and the masses of exoplanetary systems also hints at a possible underestimation of  protoplanetary disc masses  \citep{2014MNRAS.445.3315N,2018A&A...618L...3M}.

\cite{2018A&A...618L...3M} proposed two solutions to the discrepancy between observed disc masses and exoplanet system masses. The first solution was that protoplanetary discs are continuously replenished with dust, meaning that the total mass present in the disc over time is larger than the observed masses. However, it is difficult to imagine dust replenishment without replenishing gas at the same time. Since removing gas is less efficient than removing dust, additional replenishment of gas would likely lead to exceedingly long gas disc lifetimes. The second suggestion by \cite{2018A&A...618L...3M} is that planets form early, when discs are more massive. This solution seems more plausible given the large dust masses we find in young discs.

In Lupus and Taurus, the dust disc masses range between 0.1 and 100 $M_\Earth$. At 2 Myr,   the majority of the dust discs that we find are within or close to this range, but the most massive dust disc we find is $\approx\! 1000$ $M_\Earth$. This is significantly higher than in Lupus and Taurus, implying that our model overestimates the angular momentum budget of some of the cloud cores because high-angular momentum budgets lead to larger and more massive discs. It has been suggested that magnetised cloud cores regulate the angular momentum budget such that even with quite different initial conditions, the protoplanetary discs form with similar physical conditions, such as their mass \citep{2020A&A...635A..67H}. If this is the case, cloud cores with high angular momenta do not necessarily produce discs with equivalently large angular momenta. This could possibly prevent the formation of these very high-mass discs.

An alternative explanation for why we find such very high-mass discs could be that a yet-to-be found correlation exists between disc mass and formation time such that stars with large discs typically form before those with smaller discs. Indeed, observations of very young protoplanetary discs report the presence of dust discs as massive as 1000 $M_\Earth$ \citep{2018ApJS..238...19T}, suggesting that our discs that form with 1000 $M_\Earth$ are not unreasonable. However, it is possible that such discs only form in the early stages of the cluster lifetime and are then significantly depleted of dust by the time the cluster becomes comparable in age to  Lupus and Taurus. In contrast, in our model the formation time was randomly drawn between 0 and 1 Myr, allowing large discs to from relatively late.

One could also speculate that particle growth could be more efficient in these more massive and larger discs,  which is possible because the growth of particles in the outer dilute regions of these discs is not well understood. More efficient growth creates larger dust particles that drift faster, resulting in a more rapid depletion of the dust disc. However, such a process would have to apply only to the largest discs.

\subsection{Observed disc radii}

In the literature, the outer radius of the dust disc is often calculated as the radius containing some percentage of the total observed flux density, for example 68\% or 95\%  \citep[see e.g.][]{2018ApJ...869L..42H,2019A&A...626L...2F}. Under the assumption of an optically thin dust disc, the flux density is directly proportional to dust mass as
\begin{align}
F_\nu = \dfrac{B\left(\overline{T}_\mathrm{d}\right)\kappa_\nu}{D^2}M_\mathrm{d},
\end{align}
where $B\left(T_\mathrm{d}\right)$ is the Planck function at the average dust temperature $T_\mathrm{d}$, $\kappa_\nu$ is the frequency dependent opacity, and $D$ is the distance to the source. The disc temperature is often taken as $T_\mathrm{d} = 20$ \citep[see e.g.][]{2016ApJ...828...46A}, although models where $T_\mathrm{d}$ depends on the stellar luminosity have also been used \citep{2016ApJ...831..125P}. In this work we simply define the disc radius as the distance from the central star that contains 95\% of the total mass. 

We first look at the  reference case (Sect. \ref{sec:results:ref}). Figure 9 shows the stellar gas accretion rate as a function of the dust disc radius. This relation shows a similar behaviour to the stellar gas accretion rate versus dust disc mass shown in Fig. \ref{Fig:Nom:MdotvsMdisc}, but with a few notable differences. During the formation phase there are two regimes. One during the first half of the formation phase where both the accretion rate and the dust disc radius increase as the disc builds up, and one in the second half where the radius increases but the accretion rate remains almost constant. 
After the formation phase is over, the accretion rate begins to decrease, while the dust disc radius still increases because it is strongly coupled to the viscously evolving gas. At $t\approx\! 7.5\times 10^5$ yr, the dust in the outer parts of the disc begins to drift rapidly and the disc radius starts to decrease. Comparing this to Fig. \ref{Fig:Nom:MdotvsMdisc}, we can see that the dust drift turn-over takes place earlier when considering the dust disc radius instead of the dust disc mass. There is a lag between when dust starts drifting rapidly in the outer portions of the disc and when the mass of the disc starts to decrease. 

\begin{figure}[t]
   \resizebox{\hsize}{!}{\includegraphics[width=\hsize]{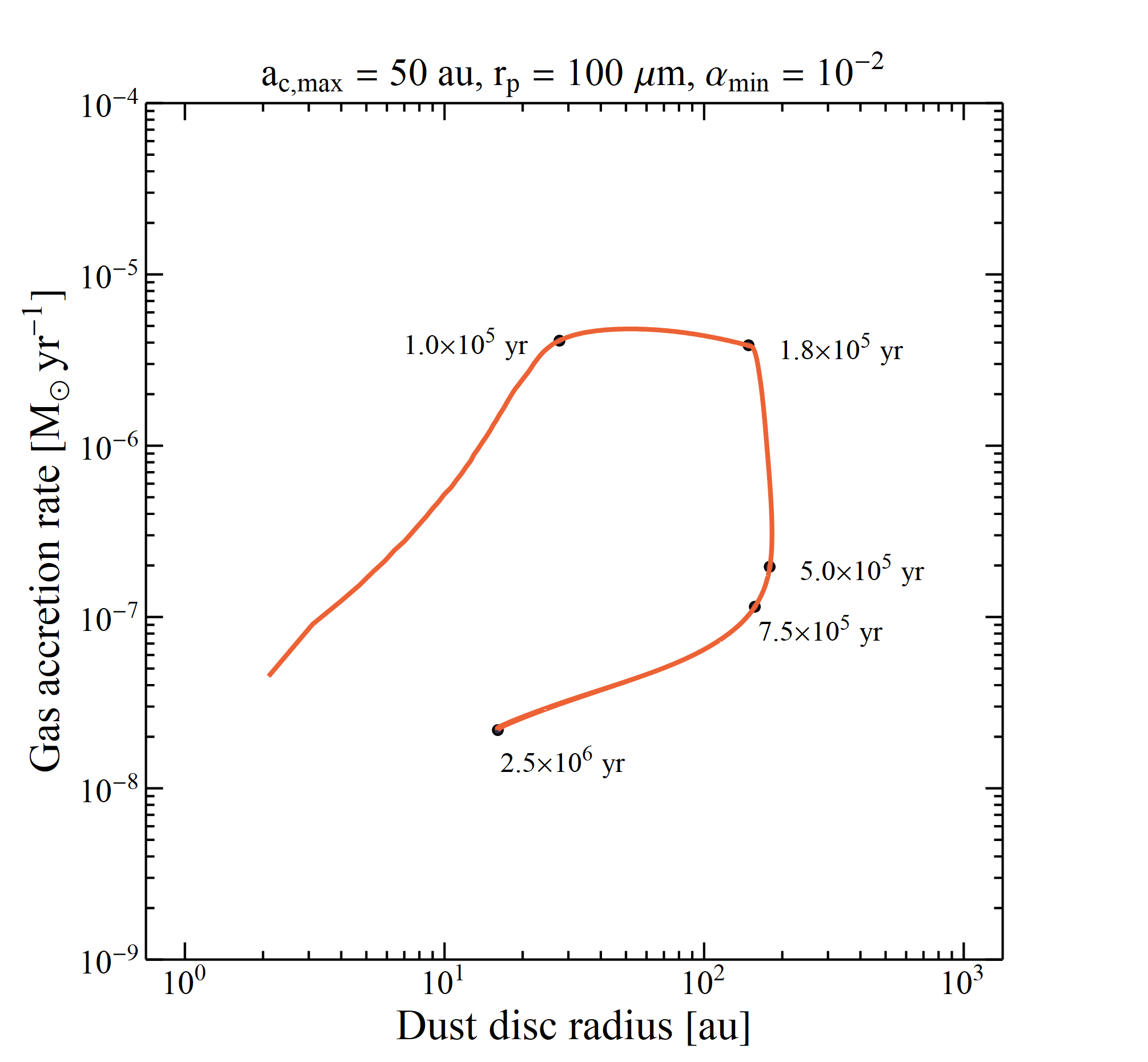} }
      \caption{  Gas accretion rate as a function of the dust disc radius for the  reference case.   During the formation phase there is a first regime where the disc radius and accretion rate both increase. In the latter half of the formation phase the disc radius continues to increase, while the gas accretion rate remains almost constant. Once the formation phase is over the accretion rate drops while the dust disc radius still continues to increase. At $t\approx\! 7.5\times 10^5$ yr the disc radius starts to decrease as the outer parts of the disc begin to drift rapidly. After $t\approx\! 2.5\times 10^6$ yr the disc is  drained of dust.}
         \label{Fig:Nom:MdvsRd}
   \end{figure}

\begin{figure}[t]
   \resizebox{\hsize}{!}{\includegraphics[width=\hsize]{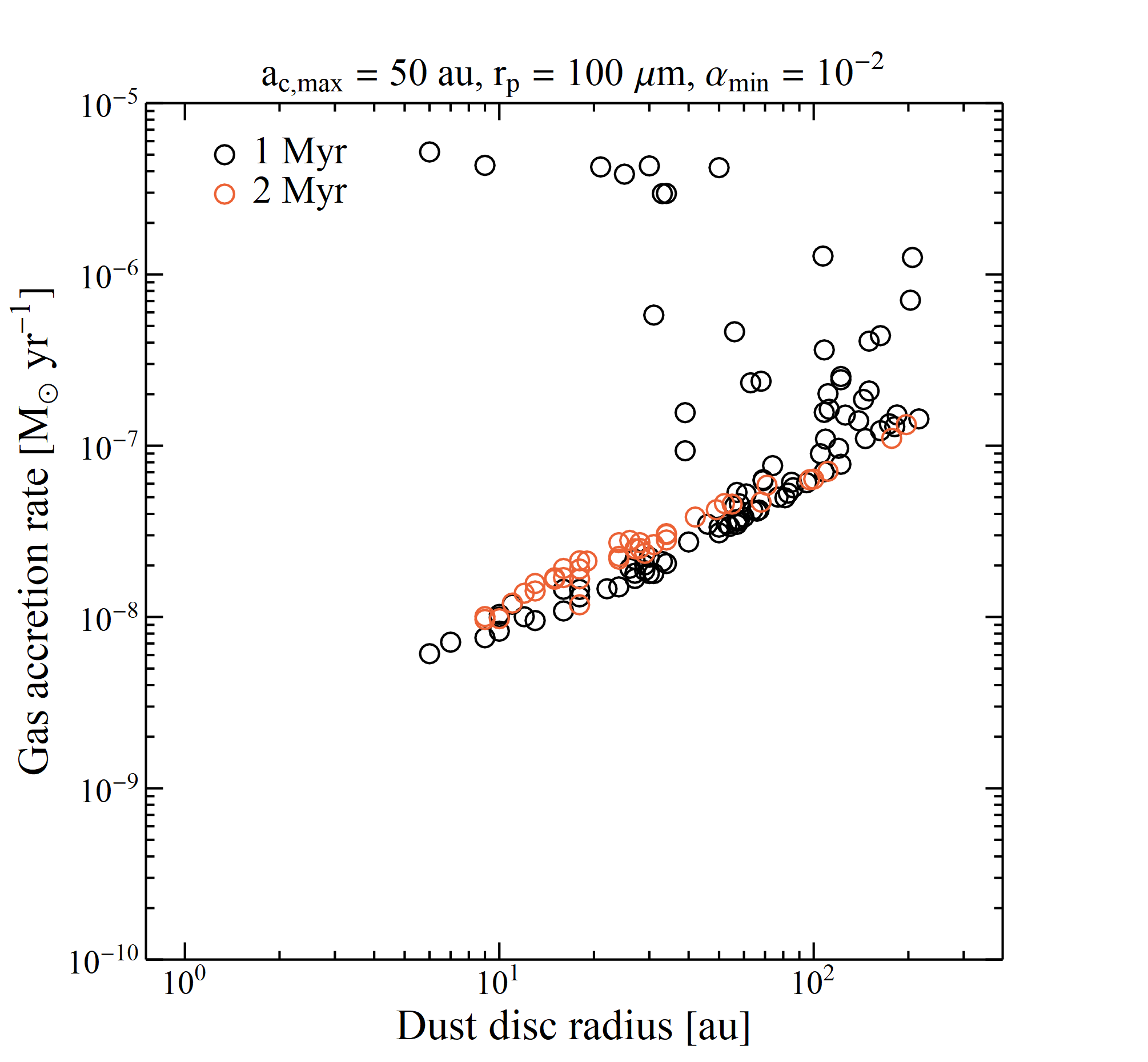} }
      \caption{  Gas accretion rate as a function of the dust disc radius of the cluster of discs at 1 Myr (black) and 2 Myr (red). Larger disc radii generally correspond to higher accretion rates.  As the discs evolve, both disc radii and accretion rate decrease.}
         \label{Fig:Nom:MdvsRdCluster}
   \end{figure}

We can also look at the stellar gas accretion rate versus dust disc radius for the cluster of stars. This is shown in Fig. \ref{Fig:Nom:MdvsRdCluster}. The black circles show the cluster at 1 Myr and the red circles show the cluster at 2 Myr. In general, a larger disc radius corresponds to a higher gas accretion rate. This is because larger discs typically come from more massive cloud cores. These cores form more massive discs with higher gas surface densities, which leads to higher gas accretion rates.  Discs with very high accretion rates are still in their formation phase.

This range of  values for dust disc radius line up well with observational measurements of the Lupus and Taurus  clusters, that place dust disc radii between a few  to a few hundred astronomical units \citep{2018ApJ...859...21A,2019ApJ...882...49L}.  

\begin{figure}[]
\resizebox{\hsize}{!}{\includegraphics[width=\hsize]{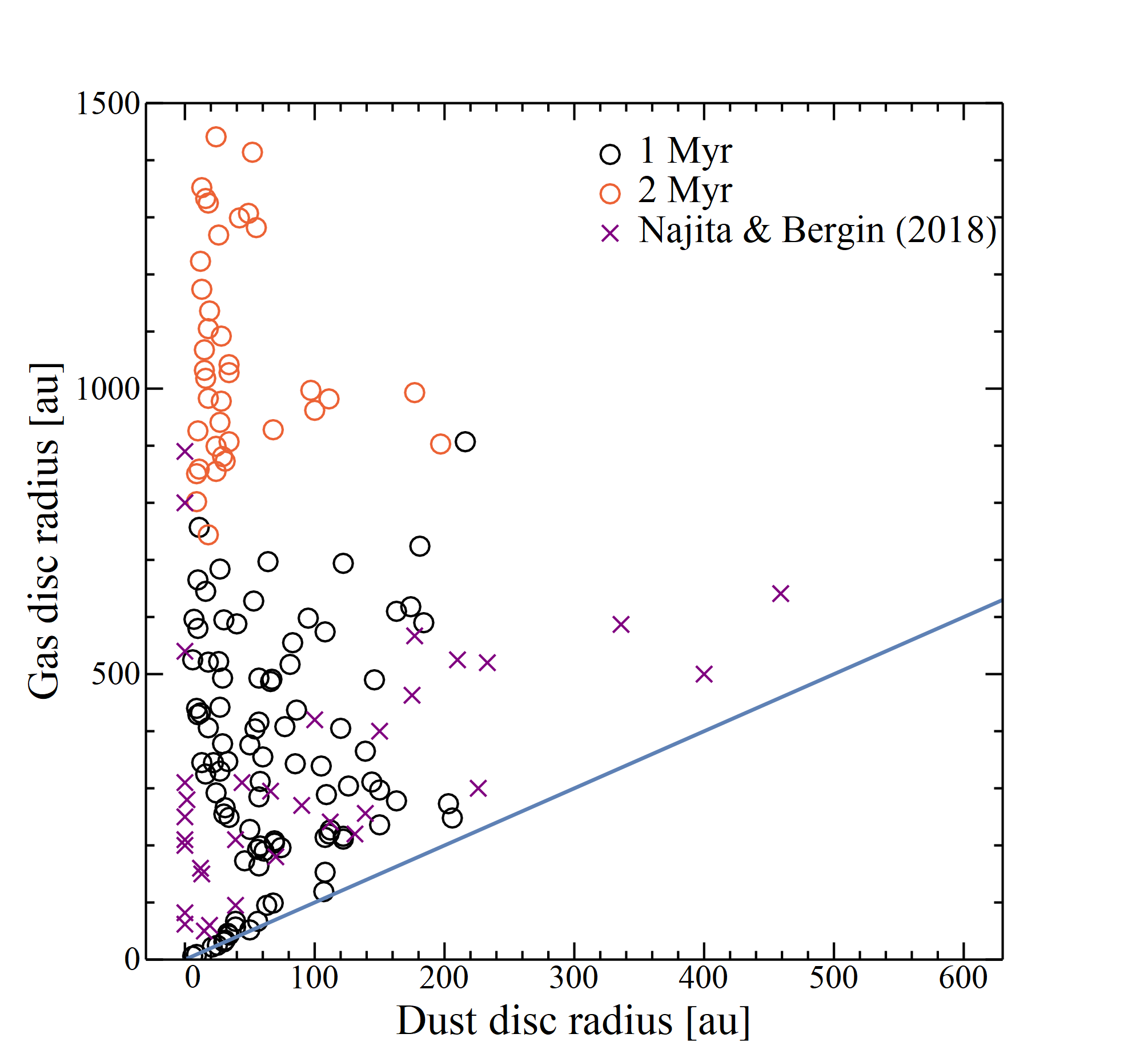} }
      \caption{  Gas disc radius as a function of dust disc radius for the cluster of discs at 1 Myr (black circles) and 2 Myr (red circles). The purple crosses are observational measurements taken from \cite{2018ApJ...864..168N}. The blue line shows $R_  \mathrm{g} = R_  \mathrm{d}$. At 1 Myr our results agree well with observations, but at 2 Myr the gas disc radii that we find are a factor of two to three higher than observations.
      }
         \label{Fig:app:RgvsRd}
   \end{figure}

For a given disc, observations show that the gas disc has a radius that is typically 1.5-8 times larger than the dust disc radius \citep{2018ApJ...864..168N}. Figure \ref{Fig:app:RgvsRd} shows the gas disc radius as a function of the dust disc radius for the modelled cluster at 1 and 2 Myr in black and red circles respectively, and
the data from \cite{2018ApJ...864..168N} are shown as purple crosses. At a cluster age of 1 Myr our results appear to agree well with these observations. These observed discs, drawn from different star-forming regions, have ages that have been estimated to be between 0.6 and 20.9 Myr.  We show in Fig 11 our results at 1 and 2 Myr. At a cluster age of 1 Myr our results appear to agree well with these observations.
However, at 2 Myr the gas disc radii in our cluster are all larger than the observations by a factor of a few. One explanation for why we find much larger discs is that we do not include external photoevaporation in our model, which can play a significant role in truncating or clearing gas discs entirely \citep[e.g.][]{2018MNRAS.478.2700W,2019MNRAS.485.4893N,2019MNRAS.490.5678C}. The very large disc radii we find can also indicate, as the very high disc masses we find also do, that the angular momentum budget of the cloud cores in our model are overestimated, since this leads to a larger centrifugal radius. During the collapse of magnetised cloud cores it is possible that the angular momentum is regulated such that discs form with similar initial radii, which could prevent the formation of discs with very large centrifugal radii \citep{2020A&A...635A..67H}. An additional possibility is that the viscous expansion of our discs is too strong.

\subsection{Stellar gas accretion rates}

In this study we find that it is difficult to reach low accretion rates ($\lesssim 10^{-9}$ $M_\sun$/yr) within  a few million years for a large range of disc masses and centrifugal radii; such low accretion rates are nevertheless often observed \citep[see e.g.][]{2016A&A...591L...3M}. One way to attempt to remedy this is to decrease the value of the viscosity parameter $\alpha_\mathrm{min}$, because this latter directly sets the gas accretion rate for a given surface density of gas. However, decreasing $\alpha_\mathrm{min}$ also decreases the speed of the viscous spreading, which in turn keeps the gas surface density higher for longer, increasing the gas accretion rate. Decreasing $\alpha_\mathrm{min}$ by an order of magnitude to  $\alpha_\mathrm{min} = 10^{-3}$ in the reference case, we find that the gas accretion rate at the end of the 5 Myr simulation is nearly the same as when $\alpha_\mathrm{min} = 10^{-2}$, but that the gas disc is roughly twice as massive. This is discussed in more detail in Appendix \ref{app:alpha}.

If some of the accreting gas near the star is removed from the inner disc through some other physical mechanism, such as for example photoevaporation of the disc or disc winds, the measured stellar gas accretion rate  drops. Indeed, work by \cite{2009ApJ...704..989A} and \cite{2016MNRAS.461.2257K} demonstrates that photoevaporation of the disc can already be important after the first million years of disc evolution when stellar gas accretion rates are in the range $10^{-9}-10^{-8}\ M_\sun \ \mathrm{yr}^{-1}$. The mechanism removing gas would have to be efficient enough to remove enough mass such that the accretion rate drops by an order of magnitude or more, down to less than $10^{-10}$ $M_\sun$ yr$^{-1}$ , as seen for example in \cite{2016A&A...591L...3M}.

Considering the stellar gas accretion rates and gas masses found in the evolution of the stellar cluster, we generally find gas disc lifetimes that are very long. Crudely estimating the remaining gas accretion timescale as the ratio between the gas accretion rate and the gas disc mass at 3 Myr yields remaining lifetimes of between 3 and 6 Myr in the cluster of discs.  However, since the gas accretion rate continues to decrease,  the remaining lifetime of the disc is underestimated. Our model also lacks gas dissipation channels other than accretion onto the star. Such methods of dissipating gas from the disc include photoevaporation \citep{2010MNRAS.401.1415O,2011MNRAS.412...13O,2012MNRAS.422.1880O} and accretion onto planets. If photoevaporation were to occur at a rate of around $10^{-9}$ $M_\sun$ yr$^{-1}$ then this process would begin to be important in the disc with the lowest gas masses that we find at 3 Myr, reducing the remaining lifetime. In a model similar to ours, but including photoevaporation,  typical gas disc lifetimes between 2 and 4 Myr have been found \citep{2016MNRAS.461.2257K}.

\subsection{Dust pile up}

\begin{figure}[t]
   \resizebox{\hsize}{!}{\includegraphics[width=\hsize]{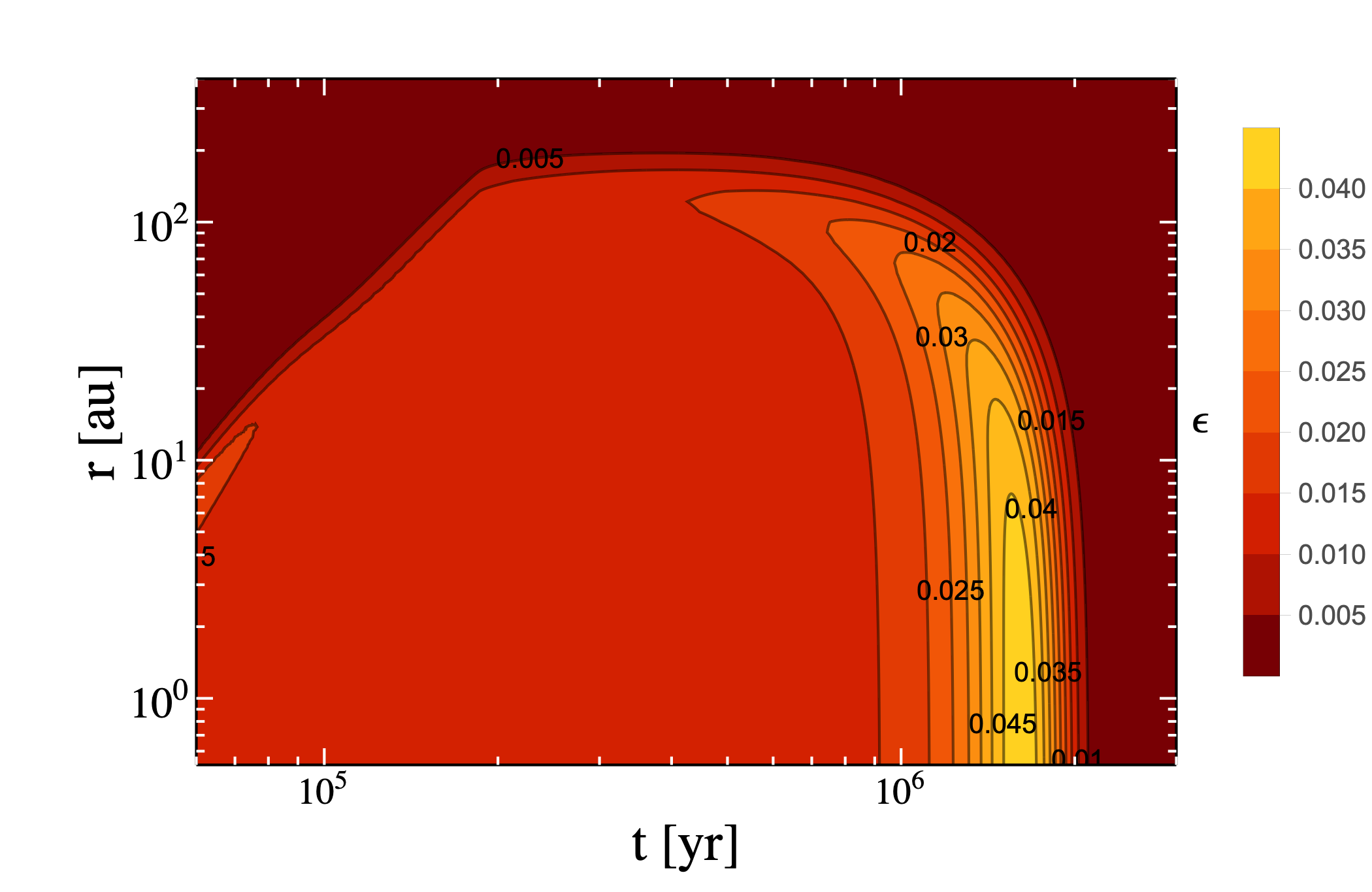} }
      \caption{ Contour plot of the dust-to-gas ratio over time and disc radius for the  reference case. For the first $\approx 0.8$ Myr the dust-to-gas ratio remains at the initial ISM value of 1/100. After this the dust-to-gas ratio begins to increase in the inner 100 au of the disc due to the pile up of dust caused by rapid radial dust drift. The increase is strongest in the inner $\approx\! 5$ au. The pile up occurs because dust drift is faster in the outer parts of the disc compared to the inner parts.
      }
         \label{Fig:Nom:DTG}
   \end{figure}

During the rapid dust drift phase we find that dust piles up, increasing the  local  dust-to-gas ratio defined as
\begin{align}
\epsilon = \Sigma_\mathrm{d}/\Sigma_\mathrm{g},
\end{align} 
above the initial value of $\epsilon_\mathrm{in} = 0.01$ at orbital radii out to 100 au.
 The evolution of the dust-to-gas ratio is shown in Fig. \ref{Fig:Nom:DTG}, which is a contour plot where the colour indicates the dust-to-gas ratio as a function of time and orbital radius of the  reference case. We can see that in the range $\approx \! 1-2$ Myr, the dust-to-gas ratio increases by a factor of several in the inner 10 au of the disc, and by a factor of two out to almost 100 au. In the inner few astronomical units we reach dust-to-gas ratios as high as 0.045. 
 
 This pile up of dust occurs because drift rates for fixed-radius particles are such that rapid drift begins furthest out in the disc. This leads to an increase in the dust surface density within this inwards-expanding front of rapid dust drift. A similar pile-up for constant-sized dust was also reported by \cite{2002ApJ...580..494Y}. The pile up is driven by the same physical mechanism in both models. However, our models differ, because we use an evolving gas disc, which calculates  the outer disc radius self-consistently and allows for viscous spreading of dust.

\begin{figure}[t]
   \resizebox{\hsize}{!}{\includegraphics[width=\hsize]{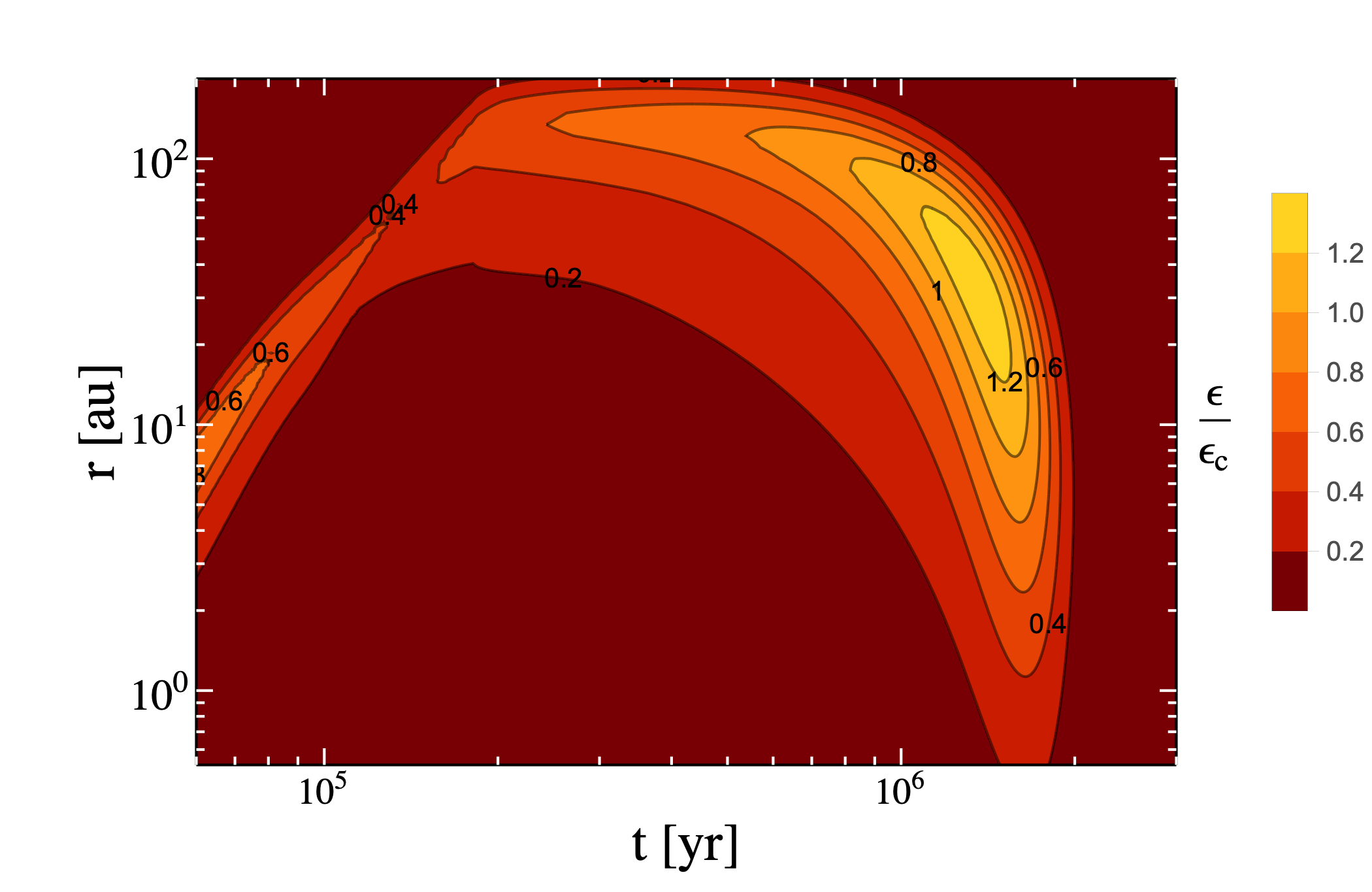} }
      \caption{Contour plot of the ratio between the dust-to-gas ratio and the critical dust-to-gas ratio needed for planetesimal formation via the streaming instability in the reference case. The dust-to-gas ratio starts to exceed the critical value at $\approx\! 100$ au at 0.8 Myr. Over the next $\approx\! 0.7$ Myr this super-critical region expands inwards to an inner edge of $\approx\!  8$ au. The region disappears as the dust is drained from the disc.
      }
         \label{Fig:Nom:DTGCrit}
   \end{figure}

The increase in dust-to-gas ratio is favourable for the formation of planetesimals.
The streaming instability path for forming planetesimals can become active if the dust-to-gas ratio is greater than some critical value \citep{2015A&A...579A..43C,2017A&A...606A..80Y}. This critical dust-to-gas ratio ranges from $\epsilon_\mathrm{c} \approx 0.035$ at $\tau_\mathrm{s} = 10^{-3}$ to a lower value of $\epsilon_\mathrm{c} \approx 0.014$ at $\tau_\mathrm{s} = 10^{-1}$, after which $\epsilon_\mathrm{c}$ increases again; see Eq. 8 and 9 of \cite{2017A&A...606A..80Y}.
Applying these criteria to the dust-to-gas ratio from the reference case seen in Fig. \ref{Fig:Nom:DTG}, we find the results shown in Fig. \ref{Fig:Nom:DTGCrit}. Favourable conditions for streaming instability  
are present starting at orbital radii of $\approx\! 100$ au at $\approx \! 0.8$  Myr. This streaming instability-favourable region then expands inwards to $\approx\! 8$ au at $\approx\! 1.5$ Myr. After the region reaches the innermost orbital radius it quickly disappears as the dust is drained entirely from the disc. Therefore, after 1 Myr of evolution the assumption of a canonical dust-to-gas ratio of 0.01 no longer holds.

         We can also analyse whether we have conditions for the continued growth of planetary embryos into full-sized planets. To investigate this we carry out a simple test where we place planetary embryos of mass $M = 0.1$  $M_\Earth$ at orbital radii and times where we have streaming instability conditions. We then let these cores grow via pebble accretion in the 2D Hill accretion regime \citep{2014A&A...572A.107L}. Planet migration is not included in this test.  One core is placed at an orbital radius of $r=30$ au at $t = 1.2$ Myr and the other at an orbital radius of $r =  10$ au at $t = 1.5$ Myr. We find that the core at $r = 30$ au grows to 10 $M_\Earth$ at $t \approx 1.6$ Myr. 
At $t \approx 2$ Myr it has grown to 17 $M_\Earth$. After this point the dust outside of the planets orbital radius is nearly completely drained and growth is halted. 
The core at $r = 10$ au is able to reach a mass of 3.2 $M_\Earth$ before the dust disc is drained. From the time the planet started growing, it takes $\approx 0.5$ Myr for the dust to drain at this orbital radius.

\cite{2019MNRAS.484.1574T} explored the possibility of planets growing by pebble accretion during the early Class 0 and Class 1I YSO phase, which lasts about 0.5 Myr. These authors found that a planet embryo starting at an orbital radius of 10 au is able to grow to a mass of 10 $M_\Earth$ during this early phase. While we do not find conditions for planetesimal formation this early in
the reference case, which we take as a prerequisite for forming the planetary embryos,  we find in Appendix \ref{app:dustsize} that using dust with a size of 1 mm gives conditions that are favourable of   streaming instability already at 0.35 Myr.  These findings, combined with those of \cite{2019MNRAS.484.1574T} that planet formation could occur quickly during this phase and with the observed difference in dust disc masses and exoplanetary system masses of \cite{2014MNRAS.445.3315N} and \cite{2018A&A...618L...3M}, suggest that planet formation is rapid and possibly already takes place during the early stages of disc lifetimes, when dust discs are massive. Signs of grain growth to mm sizes already in discs of $\approx 10^5$ yr old have also been found, suggesting that planet formation takes place early in the lives of protoplanetary discs  \citep{2018NatAs...2..646H}.

\subsection{Metallicity}

In our model, we assume that the initial dust-to-gas ratio (metallicity) of the cloud core is $\epsilon  = 0.01$ in order to match ISM values. This value is also quite close to the solar mass fraction of metals $Z_\odot = 0.0134$ \citep{2009ARA&A..47..481A}. 

In Appendix \ref{app:metallicity} we look at the effects of varying the initial dust-to-gas ratio in our model, using the metallicity in the solar neighbourhood and in nearby star-forming regions to set the distribution of the initial dust-to-gas ratio.
Variations in this initial dust-to-gas ratio will have some effects, such as changing the initial dust disc mass and  the evolved dust-to-gas ratio. This will in turn affect when streaming instability conditions are reached, with higher initial dust-to-gas ratios resulting in these conditions being met more easily. We find that introducing a scatter in the initial dust-to-gas ratio that is consistent with the scatter of metallicities in nearby star-forming regions introduces a small scatter of the dust disc masses, but the effect is very small.

\section{Conclusions}
\label{sec:conclusions}
 
Here, we investigate the formation and evolution of protoplanetary gas and dust discs using a 1D numerical model. We performed a parameter study to explore how the centrifugal radius of the disc and the mass of the cloud core affect the disc evolution. We then used population synthesis to model the evolution of a population of discs in a forming stellar cluster for several million years. We looked into how the dust disc mass evolves over time. A study of the scaling relation between disc and stellar properties was also carried out.

We find that the evolution of the dust component in a protoplanetary disc can be separated into three distinct phases, assuming dust has an almost constant size. The first is the formation phase during which  both the stellar gas accretion rate and the disc mass increase. This phase lasts on the order of $\approx\! 10^4-$ to $\times10^5$ yr, depending on the mass of the cloud core. The second regime is the slow-dust-drift phase. In this phase the gas surface density is high, which keeps the Stokes number of the dust low and therefore radial dust drift is slow.
  This results in a very steep dependence of the gas accretion rate on the dust disc mass. The third regime begins once the gas surface density drops sufficiently that dust drift becomes rapid; at this point the dust disc mass decreases on a timescale that is much shorter than the timescale over which gas accretion decreases. This results in an almost flat scaling between gas accretion rate and dust disc mass. For dust particles of 100 $\mu$m in size, completely clearing the protoplanetary disc of dust takes  between 2 and 4 Myr, depending on the mass of the cloud core from which the disc formed and on the angular momentum budget of the cloud core.
 
  We find that when using viscous evolution alone, it is difficult to reach gas accretion rates that match observed values, namely of $\dot{M} < 10^{-9}$, for the range of cloud core angular momentum and mass we consider. This is probably due to our neglect of disc winds, photoevaporation, and gas-accreting planets. We also find that the most massive discs in our population synthesis model are significantly more massive and larger than the most massive and largest observed protoplanetary discs in star-forming regions of similar ages.  We suspect this could be due to our model overestimating the occurrence rate of protoplanetary discs with large angular momentum budgets \citep{2020A&A...635A..67H}.
However, observations of very young discs show that such very massive discs are present \citep{2018ApJS..238...19T}, indicating that the very massive discs we find are younger than the youngest disc in the star-forming regions of similar cluster ages.
  
When evolving a whole cluster of discs we find that the average dust disc masses agree well with observations, and that the low observed average dust disc mass is consistent with radial drift of pebbles. However, we also note that studies indicate that dust disc masses may be higher than previously thought \citep{2018ApJ...868...39G,2019AJ....157..144B,2018A&A...618L...3M}.

For the cluster of discs we find a near-linear relation between the stellar gas accretion rate and the gas disc mass, with $\dot{M} \propto M_\mathrm{g}^{0.9}$ between 2 and 3 Myr. We find that a simple power law does not accurately describe the dependence of the gas accretion rate on the dust disc mass. The age spread of discs in the cluster will affect how many discs are in the regime of either slow or fast radial dust drift and in the first of the two regimes the scaling is very steep while in the second it is very flat. Since both regimes are somewhat populated during the first few million years of evolution, a single power law does not fit well.

We also found scaling relations of the disc mass as a function of the stellar mass. Over time this scaling becomes steeper and takes the value $M_\mathrm{d} \propto M_\star^{1.4-4.1}$ between 1 and 2 Myr. This steepening with time qualitatively agrees with observations \citep{2016ApJ...831..125P}, but the steepening in our model is faster than observations suggest.

 We find in all our simulations that the drifting dust piles up in the inner regions of the disc. The pile up occurs when the disc is 1-2 Myr old.  During the pile up, the dust-to-gas ratio increases in the inner 100 au of the disc, reaching values four times higher than the initial value (1/100).
The dust-to-gas ratio becomes high enough that, according to the criteria from \cite{2017A&A...606A..80Y}, the streaming instability can become active at orbital radii between 10 au and 100 au, between 0.8 Myr and 1.5 Myr. 

In the present study, we use modelling of the formation and evolution of protoplanetary discs to show that for dust particles of constant size, the protoplanetary dust discs go through three distinct phases. The build up of the disc, the slow radial drift regime, and the fast radial drift regime. Using a population synthesis model we are able to show that the low observed dust disc masses in star-forming regions are consistent with the rapid radial drift of pebbles. Dust discs can be very massive when they are young and still agree with the low observed average masses at their later evolutionary stages. This implies that planets form rapidly during the early, massive phase of the dust disc evolution.

\begin{acknowledgements}
We want to thank the anonymous referee for helpful comments that helped to improve the original manuscript. We also want to thank Carlo Manara, Giovanni Rosotti and  Per Bjerkeli for their helpful discussions, and Paul McMillan for his help.
J.A. and M.L. thanks the Wallenberg Academy Fellow (grant 2017.0287).
A.J.  thanks  the  Swedish  Research  Council  (grant  2018-04867),   the   Knut   and   Alice   Wallenberg   Foundation   (grant 2017.0287)  and  the  European  Research  Council  (ERC  Consolidator  Grant 724687-PLANETESYS)   for   research   support. 

\end{acknowledgements}

\bibliographystyle{aa}
\bibliography{References}

\begin{thebibliography}{93}
\expandafter\ifx\csname natexlab\endcsname\relax\def\natexlab#1{#1}\fi

\bibitem[{{Agurto-Gangas} {et~al.}(2019){Agurto-Gangas}, {Pineda},
  {Sz{\H{u}}cs}, {Testi}, {Tazzari}, {Miotello}, {Caselli}, {Dunham},
  {Stephens}, \& {Bourke}}]{2019A&A...623A.147A}
{Agurto-Gangas}, C., {Pineda}, J.~E., {Sz{\H{u}}cs}, L., {et~al.} 2019, \aap,
  623, A147

\bibitem[{{Alcal{\'a}} {et~al.}(2017){Alcal{\'a}}, {Manara}, {Natta}, {Frasca},
  {Testi}, {Nisini}, {Stelzer}, {Williams}, {Antoniucci}, {Biazzo}, {Covino},
  {Esposito}, {Getman}, \& {Rigliaco}}]{2017A&A...600A..20A}
{Alcal{\'a}}, J.~M., {Manara}, C.~F., {Natta}, A., {et~al.} 2017, \aap, 600,
  A20

\bibitem[{{Alexander} \& {Armitage}(2009)}]{2009ApJ...704..989A}
{Alexander}, R.~D. \& {Armitage}, P.~J. 2009, \apj, 704, 989

\bibitem[{{Andrews} {et~al.}(2018){Andrews}, {Huang}, {P{\'e}rez}, {Isella},
  {Dullemond}, {Kurtovic}, {Guzm{\'a}n}, {Carpenter}, {Wilner}, {Zhang}, {Zhu},
  {Birnstiel}, {Bai}, {Benisty}, {Hughes}, {{\"O}berg}, \&
  {Ricci}}]{2018ApJ...869L..41A}
{Andrews}, S.~M., {Huang}, J., {P{\'e}rez}, L.~M., {et~al.} 2018, \apjl, 869,
  L41

\bibitem[{{Ansdell} {et~al.}(2018){Ansdell}, {Williams}, {Trapman}, {van
  Terwisga}, {Facchini}, {Manara}, {van der Marel}, {Miotello}, {Tazzari},
  {Hogerheijde}, {Guidi}, {Testi}, \& {van Dishoeck}}]{2018ApJ...859...21A}
{Ansdell}, M., {Williams}, J.~P., {Trapman}, L., {et~al.} 2018, \apj, 859, 21

\bibitem[{{Ansdell} {et~al.}(2016){Ansdell}, {Williams}, {van der Marel},
  {Carpenter}, {Guidi}, {Hogerheijde}, {Mathews}, {Manara}, {Miotello},
  {Natta}, {Oliveira}, {Tazzari}, {Testi}, {van Dishoeck}, \& {van
  Terwisga}}]{2016ApJ...828...46A}
{Ansdell}, M., {Williams}, J.~P., {van der Marel}, N., {et~al.} 2016, \apj,
  828, 46

\bibitem[{{Asplund} {et~al.}(2009){Asplund}, {Grevesse}, {Sauval}, \&
  {Scott}}]{2009ARA&A..47..481A}
{Asplund}, M., {Grevesse}, N., {Sauval}, A.~J., \& {Scott}, P. 2009, \araa, 47,
  481

\bibitem[{{Bai} \& {Stone}(2013)}]{2013ApJ...769...76B}
{Bai}, X.-N. \& {Stone}, J.~M. 2013, \apj, 769, 76

\bibitem[{{Ballering} \& {Eisner}(2019)}]{2019AJ....157..144B}
{Ballering}, N.~P. \& {Eisner}, J.~A. 2019, \aj, 157, 144

\bibitem[{{Baraffe} {et~al.}(2002){Baraffe}, {Chabrier}, {Allard}, \&
  {Hauschildt}}]{2002A&A...382..563B}
{Baraffe}, I., {Chabrier}, G., {Allard}, F., \& {Hauschildt}, P.~H. 2002, \aap,
  382, 563

\bibitem[{{Barenfeld} {et~al.}(2016){Barenfeld}, {Carpenter}, {Ricci}, \&
  {Isella}}]{2016ApJ...827..142B}
{Barenfeld}, S.~A., {Carpenter}, J.~M., {Ricci}, L., \& {Isella}, A. 2016,
  \apj, 827, 142

\bibitem[{{Bath} \& {Pringle}(1981)}]{1981MNRAS.194..967B}
{Bath}, G.~T. \& {Pringle}, J.~E. 1981, \mnras, 194, 967

\bibitem[{{Bayo} {et~al.}(2012){Bayo}, {Barrado}, {Hu{\'e}lamo},
  {Morales-Calder{\'o}n}, {Melo}, {Stauffer}, \&
  {Stelzer}}]{2012A&A...547A..80B}
{Bayo}, A., {Barrado}, D., {Hu{\'e}lamo}, N., {et~al.} 2012, \aap, 547, A80

\bibitem[{{Bergin} {et~al.}(2013){Bergin}, {Cleeves}, {Gorti}, {Zhang},
  {Blake}, {Green}, {Andrews}, {Evans}, {Henning}, {{\"O}berg}, {Pontoppidan},
  {Qi}, {Salyk}, \& {van Dishoeck}}]{2013Natur.493..644B}
{Bergin}, E.~A., {Cleeves}, L.~I., {Gorti}, U., {et~al.} 2013, \nat, 493, 644

\bibitem[{{Birnstiel} {et~al.}(2010){Birnstiel}, {Dullemond}, \&
  {Brauer}}]{2010A&A...513A..79B}
{Birnstiel}, T., {Dullemond}, C.~P., \& {Brauer}, F. 2010, \aap, 513, A79

\bibitem[{{Birnstiel} {et~al.}(2012){Birnstiel}, {Klahr}, \&
  {Ercolano}}]{2012A&A...539A.148B}
{Birnstiel}, T., {Klahr}, H., \& {Ercolano}, B. 2012, \aap, 539, A148

\bibitem[{{Boley} {et~al.}(2006){Boley}, {Mej{\'\i}a}, {Durisen}, {Cai},
  {Pickett}, \& {D'Alessio}}]{2006ApJ...651..517B}
{Boley}, A.~C., {Mej{\'\i}a}, A.~C., {Durisen}, R.~H., {et~al.} 2006, \apj,
  651, 517

\bibitem[{{Bonnor}(1956)}]{1956MNRAS.116..351B}
{Bonnor}, W.~B. 1956, \mnras, 116, 351

\bibitem[{{Booth} {et~al.}(2017){Booth}, {Clarke}, {Madhusudhan}, \&
  {Ilee}}]{2017MNRAS.469.3994B}
{Booth}, R.~A., {Clarke}, C.~J., {Madhusudhan}, N., \& {Ilee}, J.~D. 2017,
  \mnras, 469, 3994

\bibitem[{{Bragan{\c{c}}a} {et~al.}(2019){Bragan{\c{c}}a}, {Daflon}, {Lanz},
  {Cunha}, {Bensby}, {McMillan}, {Garmany}, {Glaspey}, {Borges Fernandes},
  {Oey}, \& {Hubeny}}]{2019A&A...625A.120B}
{Bragan{\c{c}}a}, G.~A., {Daflon}, S., {Lanz}, T., {et~al.} 2019, \aap, 625,
  A120

\bibitem[{{Brauer} {et~al.}(2008){Brauer}, {Dullemond}, \&
  {Henning}}]{2008A&A...480..859B}
{Brauer}, F., {Dullemond}, C.~P., \& {Henning}, T. 2008, \aap, 480, 859

\bibitem[{{Carpenter} {et~al.}(2006){Carpenter}, {Mamajek}, {Hillenbrand}, \&
  {Meyer}}]{2006ApJ...651L..49C}
{Carpenter}, J.~M., {Mamajek}, E.~E., {Hillenbrand}, L.~A., \& {Meyer}, M.~R.
  2006, \apjl, 651, L49

\bibitem[{{Carrera} {et~al.}(2015){Carrera}, {Johansen}, \&
  {Davies}}]{2015A&A...579A..43C}
{Carrera}, D., {Johansen}, A., \& {Davies}, M.~B. 2015, \aap, 579, A43

\bibitem[{{Chen} \& {Ostriker}(2018)}]{2018ApJ...865...34C}
{Chen}, C.-Y. \& {Ostriker}, E.~C. 2018, \apj, 865, 34

\bibitem[{{Cieza} {et~al.}(2016){Cieza}, {Casassus}, {Tobin}, {Bos},
  {Williams}, {Perez}, {Zhu}, {Caceres}, {Canovas}, {Dunham}, {Hales},
  {Prieto}, {Principe}, {Schreiber}, {Ruiz-Rodriguez}, \&
  {Zurlo}}]{2016Natur.535..258C}
{Cieza}, L.~A., {Casassus}, S., {Tobin}, J., {et~al.} 2016, \nat, 535, 258

\bibitem[{{Concha-Ram{\'\i}rez} {et~al.}(2019){Concha-Ram{\'\i}rez}, {Wilhelm},
  {Portegies Zwart}, \& {Haworth}}]{2019MNRAS.490.5678C}
{Concha-Ram{\'\i}rez}, F., {Wilhelm}, M. J.~C., {Portegies Zwart}, S., \&
  {Haworth}, T.~J. 2019, \mnras, 490, 5678

\bibitem[{{D'Orazi} {et~al.}(2011){D'Orazi}, {Biazzo}, \&
  {Randich}}]{2011A&A...526A.103D}
{D'Orazi}, V., {Biazzo}, K., \& {Randich}, S. 2011, \aap, 526, A103

\bibitem[{{D'Orazi} {et~al.}(2009){D'Orazi}, {Randich}, {Flaccomio}, {Palla},
  {Sacco}, \& {Pallavicini}}]{2009A&A...501..973D}
{D'Orazi}, V., {Randich}, S., {Flaccomio}, E., {et~al.} 2009, \aap, 501, 973

\bibitem[{{Dullemond} {et~al.}(2006){Dullemond}, {Natta}, \&
  {Testi}}]{2006ApJ...645L..69D}
{Dullemond}, C.~P., {Natta}, A., \& {Testi}, L. 2006, \apjl, 645, L69

\bibitem[{{Ebert}(1957)}]{1957ZA.....42..263E}
{Ebert}, R. 1957, \zap, 42, 263

\bibitem[{{Facchini} {et~al.}(2019){Facchini}, {van Dishoeck}, {Manara},
  {Tazzari}, {Maud}, {Cazzoletti}, {Rosotti}, {van der Marel}, {Pinilla}, \&
  {Clarke}}]{2019A&A...626L...2F}
{Facchini}, S., {van Dishoeck}, E.~F., {Manara}, C.~F., {et~al.} 2019, \aap,
  626, L2

\bibitem[{{Falceta-Gon{\c{c}}alves} \& {Lazarian}(2011)}]{2011ApJ...735...99F}
{Falceta-Gon{\c{c}}alves}, D. \& {Lazarian}, A. 2011, \apj, 735, 99

\bibitem[{{Fedele} {et~al.}(2010){Fedele}, {van den Ancker}, {Henning},
  {Jayawardhana}, \& {Oliveira}}]{2010A&A...510A..72F}
{Fedele}, D., {van den Ancker}, M.~E., {Henning}, T., {Jayawardhana}, R., \&
  {Oliveira}, J.~M. 2010, \aap, 510, A72

\bibitem[{{Galv{\'a}n-Madrid} {et~al.}(2018){Galv{\'a}n-Madrid}, {Liu},
  {Izquierdo}, {Miotello}, {Zhao}, {Carrasco-Gonz{\'a}lez}, {Lizano}, \&
  {Rodr{\'{\i}}guez}}]{2018ApJ...868...39G}
{Galv{\'a}n-Madrid}, R., {Liu}, H.~B., {Izquierdo}, A.~F., {et~al.} 2018, \apj,
  868, 39

\bibitem[{{Gonzalez} {et~al.}(2017){Gonzalez}, {Laibe}, \&
  {Maddison}}]{2017MNRAS.467.1984G}
{Gonzalez}, J.~F., {Laibe}, G., \& {Maddison}, S.~T. 2017, \mnras, 467, 1984

\bibitem[{{Goodman} {et~al.}(1993){Goodman}, {Benson}, {Fuller}, \&
  {Myers}}]{1993ApJ...406..528G}
{Goodman}, A.~A., {Benson}, P.~J., {Fuller}, G.~A., \& {Myers}, P.~C. 1993,
  \apj, 406, 528

\bibitem[{{Gressel} {et~al.}(2015){Gressel}, {Turner}, {Nelson}, \&
  {McNally}}]{2015ApJ...801...84G}
{Gressel}, O., {Turner}, N.~J., {Nelson}, R.~P., \& {McNally}, C.~P. 2015,
  \apj, 801, 84

\bibitem[{{G{\"u}ttler} {et~al.}(2010){G{\"u}ttler}, {Blum}, {Zsom}, {Ormel},
  \& {Dullemond}}]{2010A&A...513A..56G}
{G{\"u}ttler}, C., {Blum}, J., {Zsom}, A., {Ormel}, C.~W., \& {Dullemond},
  C.~P. 2010, \aap, 513, A56

\bibitem[{{Haisch} {et~al.}(2001){Haisch}, {Lada}, \&
  {Lada}}]{2001ApJ...553L.153H}
{Haisch}, Jr., K.~E., {Lada}, E.~A., \& {Lada}, C.~J. 2001, \apjl, 553, L153

\bibitem[{{Harsono} {et~al.}(2018){Harsono}, {Bjerkeli}, {van der Wiel},
  {Ramsey}, {Maud}, {Kristensen}, \& {J{\o}rgensen}}]{2018NatAs...2..646H}
{Harsono}, D., {Bjerkeli}, P., {van der Wiel}, M. H.~D., {et~al.} 2018, Nature
  Astronomy, 2, 646

\bibitem[{{Hayashi}(1981)}]{1981PThPS..70...35H}
{Hayashi}, C. 1981, Progress of Theoretical Physics Supplement, 70, 35

\bibitem[{{Hennebelle} {et~al.}(2020){Hennebelle}, {Commer{\c{c}}on}, {Lee}, \&
  {Charnoz}}]{2020A&A...635A..67H}
{Hennebelle}, P., {Commer{\c{c}}on}, B., {Lee}, Y.-N., \& {Charnoz}, S. 2020,
  \aap, 635, A67

\bibitem[{{Huang} {et~al.}(2018){Huang}, {Andrews}, {Dullemond}, {Isella},
  {P{\'e}rez}, {Guzm{\'a}n}, {{\"O}berg}, {Zhu}, {Zhang}, {Bai}, {Benisty},
  {Birnstiel}, {Carpenter}, {Hughes}, {Ricci}, {Weaver}, \&
  {Wilner}}]{2018ApJ...869L..42H}
{Huang}, J., {Andrews}, S.~M., {Dullemond}, C.~P., {et~al.} 2018, \apjl, 869,
  L42

\bibitem[{{Hueso} \& {Guillot}(2005)}]{2005A&A...442..703H}
{Hueso}, R. \& {Guillot}, T. 2005, \aap, 442, 703

\bibitem[{{Kataoka} {et~al.}(2016){Kataoka}, {Tsukagoshi}, {Momose}, {Nagai},
  {Muto}, {Dullemond}, {Pohl}, {Fukagawa}, {Shibai}, {Hanawa}, \&
  {Murakawa}}]{2016ApJ...831L..12K}
{Kataoka}, A., {Tsukagoshi}, T., {Momose}, M., {et~al.} 2016, \apjl, 831, L12

\bibitem[{{Kimura} {et~al.}(2016){Kimura}, {Kunitomo}, \&
  {Takahashi}}]{2016MNRAS.461.2257K}
{Kimura}, S.~S., {Kunitomo}, M., \& {Takahashi}, S.~Z. 2016, \mnras, 461, 2257

\bibitem[{{Kroupa}(2001)}]{2001MNRAS.322..231K}
{Kroupa}, P. 2001, \mnras, 322, 231

\bibitem[{{Lambrechts} \& {Johansen}(2014)}]{2014A&A...572A.107L}
{Lambrechts}, M. \& {Johansen}, A. 2014, \aap, 572, A107

\bibitem[{{Lodato} {et~al.}(2017){Lodato}, {Scardoni}, {Manara}, \&
  {Testi}}]{2017MNRAS.472.4700L}
{Lodato}, G., {Scardoni}, C.~E., {Manara}, C.~F., \& {Testi}, L. 2017, \mnras,
  472, 4700

\bibitem[{{Long} {et~al.}(2019){Long}, {Herczeg}, {Harsono}, {Pinilla},
  {Tazzari}, {Manara}, {Pascucci}, {Cabrit}, {Nisini}, {Johnstone}, {Edwards},
  {Salyk}, {Menard}, {Lodato}, {Boehler}, {Mace}, {Liu}, {Mulders}, {Hendler},
  {Ragusa}, {Fischer}, {Banzatti}, {Rigliaco}, {van de Plas}, {Dipierro},
  {Gully-Santiago}, \& {Lopez-Valdivia}}]{2019ApJ...882...49L}
{Long}, F., {Herczeg}, G.~J., {Harsono}, D., {et~al.} 2019, \apj, 882, 49

\bibitem[{{Manara} {et~al.}(2016{\natexlab{a}}){Manara}, {Fedele}, {Herczeg},
  \& {Teixeira}}]{2016A&A...585A.136M}
{Manara}, C.~F., {Fedele}, D., {Herczeg}, G.~J., \& {Teixeira}, P.~S.
  2016{\natexlab{a}}, \aap, 585, A136

\bibitem[{{Manara} {et~al.}(2018){Manara}, {Morbidelli}, \&
  {Guillot}}]{2018A&A...618L...3M}
{Manara}, C.~F., {Morbidelli}, A., \& {Guillot}, T. 2018, \aap, 618, L3

\bibitem[{{Manara} {et~al.}(2016{\natexlab{b}}){Manara}, {Rosotti}, {Testi},
  {Natta}, {Alcal{\'a}}, {Williams}, {Ansdell}, {Miotello}, {van der Marel},
  {Tazzari}, {Carpenter}, {Guidi}, {Mathews}, {Oliveira}, {Prusti}, \& {van
  Dishoeck}}]{2016A&A...591L...3M}
{Manara}, C.~F., {Rosotti}, G., {Testi}, L., {et~al.} 2016{\natexlab{b}}, \aap,
  591, L3

\bibitem[{{Manara} {et~al.}(2017){Manara}, {Testi}, {Herczeg}, {Pascucci},
  {Alcal{\'a}}, {Natta}, {Antoniucci}, {Fedele}, {Mulders}, {Henning},
  {Mohanty}, {Prusti}, \& {Rigliaco}}]{2017A&A...604A.127M}
{Manara}, C.~F., {Testi}, L., {Herczeg}, G.~J., {et~al.} 2017, \aap, 604, A127

\bibitem[{{McClure} {et~al.}(2016){McClure}, {Bergin}, {Cleeves}, {van
  Dishoeck}, {Blake}, {Evans}, {Green}, {Henning}, {{\"O}berg}, {Pontoppidan},
  \& {Salyk}}]{2016ApJ...831..167M}
{McClure}, M.~K., {Bergin}, E.~A., {Cleeves}, L.~I., {et~al.} 2016, \apj, 831,
  167

\bibitem[{{McKee} \& {Ostriker}(2007)}]{2007ARA&A..45..565M}
{McKee}, C.~F. \& {Ostriker}, E.~C. 2007, \araa, 45, 565

\bibitem[{{Miotello} {et~al.}(2014){Miotello}, {Testi}, {Lodato}, {Ricci},
  {Rosotti}, {Brooks}, {Maury}, \& {Natta}}]{2014A&A...567A..32M}
{Miotello}, A., {Testi}, L., {Lodato}, G., {et~al.} 2014, \aap, 567, A32

\bibitem[{{Miotello} {et~al.}(2017){Miotello}, {van Dishoeck}, {Williams},
  {Ansdell}, {Guidi}, {Hogerheijde}, {Manara}, {Tazzari}, {Testi}, {van der
  Marel}, \& {van Terwisga}}]{2017A&A...599A.113M}
{Miotello}, A., {van Dishoeck}, E.~F., {Williams}, J.~P., {et~al.} 2017, \aap,
  599, A113

\bibitem[{{Mori} {et~al.}(2019){Mori}, {Bai}, \&
  {Okuzumi}}]{2019ApJ...872...98M}
{Mori}, S., {Bai}, X.-N., \& {Okuzumi}, S. 2019, \apj, 872, 98

\bibitem[{{Musiolik} \& {Wurm}(2019)}]{2019ApJ...873...58M}
{Musiolik}, G. \& {Wurm}, G. 2019, \apj, 873, 58

\bibitem[{{Muzerolle} {et~al.}(2003){Muzerolle}, {Hillenbrand}, {Calvet},
  {Brice{\~n}o}, \& {Hartmann}}]{2003ApJ...592..266M}
{Muzerolle}, J., {Hillenbrand}, L., {Calvet}, N., {Brice{\~n}o}, C., \&
  {Hartmann}, L. 2003, \apj, 592, 266

\bibitem[{{Najita} \& {Bergin}(2018)}]{2018ApJ...864..168N}
{Najita}, J.~R. \& {Bergin}, E.~A. 2018, \apj, 864, 168

\bibitem[{{Najita} \& {Kenyon}(2014)}]{2014MNRAS.445.3315N}
{Najita}, J.~R. \& {Kenyon}, S.~J. 2014, \mnras, 445, 3315

\bibitem[{{Nakagawa} {et~al.}(1986){Nakagawa}, {Sekiya}, \&
  {Hayashi}}]{1986Icar...67..375N}
{Nakagawa}, Y., {Sekiya}, M., \& {Hayashi}, C. 1986, \icarus, 67, 375

\bibitem[{{Natta} {et~al.}(2004){Natta}, {Testi}, {Muzerolle}, {Randich},
  {Comer{\'o}n}, \& {Persi}}]{2004A&A...424..603N}
{Natta}, A., {Testi}, L., {Muzerolle}, J., {et~al.} 2004, \aap, 424, 603

\bibitem[{{Nicholson} {et~al.}(2019){Nicholson}, {Parker}, {Church}, {Davies},
  {Fearon}, \& {Walton}}]{2019MNRAS.485.4893N}
{Nicholson}, R.~B., {Parker}, R.~J., {Church}, R.~P., {et~al.} 2019, \mnras,
  485, 4893

\bibitem[{{Nieva} \& {Przybilla}(2008)}]{2008A&A...481..199N}
{Nieva}, M.~F. \& {Przybilla}, N. 2008, \aap, 481, 199

\bibitem[{{Ohashi} {et~al.}(2018){Ohashi}, {Kataoka}, {Nagai}, {Momose},
  {Muto}, {Hanawa}, {Fukagawa}, {Tsukagoshi}, {Murakawa}, \&
  {Shibai}}]{2018ApJ...864...81O}
{Ohashi}, S., {Kataoka}, A., {Nagai}, H., {et~al.} 2018, \apj, 864, 81

\bibitem[{{Owen} {et~al.}(2012){Owen}, {Clarke}, \&
  {Ercolano}}]{2012MNRAS.422.1880O}
{Owen}, J.~E., {Clarke}, C.~J., \& {Ercolano}, B. 2012, \mnras, 422, 1880

\bibitem[{{Owen} {et~al.}(2011){Owen}, {Ercolano}, \&
  {Clarke}}]{2011MNRAS.412...13O}
{Owen}, J.~E., {Ercolano}, B., \& {Clarke}, C.~J. 2011, \mnras, 412, 13

\bibitem[{{Owen} {et~al.}(2010){Owen}, {Ercolano}, {Clarke}, \& {Alexand
  er}}]{2010MNRAS.401.1415O}
{Owen}, J.~E., {Ercolano}, B., {Clarke}, C.~J., \& {Alexand er}, R.~D. 2010,
  \mnras, 401, 1415

\bibitem[{{Pascucci} {et~al.}(2016){Pascucci}, {Testi}, {Herczeg}, {Long},
  {Manara}, {Hendler}, {Mulders}, {Krijt}, {Ciesla}, \&
  {Henning}}]{2016ApJ...831..125P}
{Pascucci}, I., {Testi}, L., {Herczeg}, G.~J., {et~al.} 2016, \apj, 831, 125

\bibitem[{{Pinte} {et~al.}(2016){Pinte}, {Dent}, {M{\'e}nard}, {Hales}, {Hill},
  {Cortes}, \& {de Gregorio-Monsalvo}}]{2016ApJ...816...25P}
{Pinte}, C., {Dent}, W.~R.~F., {M{\'e}nard}, F., {et~al.} 2016, \apj, 816, 25

\bibitem[{{Pringle}(1981)}]{1981ARA&A..19..137P}
{Pringle}, J.~E. 1981, \araa, 19, 137

\bibitem[{{Sakai} {et~al.}(2016){Sakai}, {Oya}, {L{\'o}pez-Sepulcre},
  {Watanabe}, {Sakai}, {Hirota}, {Aikawa}, {Ceccarelli}, {Lefloch}, {Caux},
  {Vastel}, {Kahane}, \& {Yamamoto}}]{2016ApJ...820L..34S}
{Sakai}, N., {Oya}, Y., {L{\'o}pez-Sepulcre}, A., {et~al.} 2016, \apjl, 820,
  L34

\bibitem[{{Sakai} {et~al.}(2014){Sakai}, {Oya}, {Sakai}, {Watanabe}, {Hirota},
  {Ceccarelli}, {Kahane}, {Lopez-Sepulcre}, {Lefloch}, {Vastel}, {Bottinelli},
  {Caux}, {Coutens}, {Aikawa}, {Takakuwa}, {Ohashi}, {Yen}, \&
  {Yamamoto}}]{2014ApJ...791L..38S}
{Sakai}, N., {Oya}, Y., {Sakai}, T., {et~al.} 2014, \apjl, 791, L38

\bibitem[{{Sch{\"o}nrich} \& {Binney}(2009)}]{2009MNRAS.399.1145S}
{Sch{\"o}nrich}, R. \& {Binney}, J. 2009, \mnras, 399, 1145

\bibitem[{{Shakura} \& {Sunyaev}(1973)}]{1973A&A....24..337S}
{Shakura}, N.~I. \& {Sunyaev}, R.~A. 1973, \aap, 24, 337

\bibitem[{{Shu}(1977)}]{1977ApJ...214..488S}
{Shu}, F.~H. 1977, \apj, 214, 488

\bibitem[{{Spina} {et~al.}(2014){Spina}, {Randich}, {Palla}, {Biazzo}, {Sacco},
  {Alfaro}, {Franciosini}, {Magrini}, {Morbidelli}, {Frasca}, {Adibekyan},
  {Delgado-Mena}, {Sousa}, {Gonz{\'a}lez Hern{\'a}ndez}, {Montes}, {Tabernero},
  {Tautvai{\v{s}}ien{\.{e}}}, {Bonito}, {Lanzafame}, {Gilmore}, {Jeffries},
  {Vallenari}, {Bensby}, {Bragaglia}, {Flaccomio}, {Korn}, {Pancino},
  {Recio-Blanco}, {Smiljanic}, {Bergemann}, {Costado}, {Damiani}, {Hill},
  {Hourihane}, {Jofr{\'e}}, {de Laverny}, {Lardo}, {Masseron}, {Prisinzano}, \&
  {Worley}}]{2014A&A...568A...2S}
{Spina}, L., {Randich}, S., {Palla}, F., {et~al.} 2014, \aap, 568, A2

\bibitem[{{Takahashi} {et~al.}(2013){Takahashi}, {Inutsuka}, \&
  {Machida}}]{2013ApJ...770...71T}
{Takahashi}, S.~Z., {Inutsuka}, S.-i., \& {Machida}, M.~N. 2013, \apj, 770, 71

\bibitem[{{Takeuchi} {et~al.}(2005){Takeuchi}, {Clarke}, \&
  {Lin}}]{2005ApJ...627..286T}
{Takeuchi}, T., {Clarke}, C.~J., \& {Lin}, D.~N.~C. 2005, \apj, 627, 286

\bibitem[{{Takeuchi} \& {Lin}(2005)}]{2005ApJ...623..482T}
{Takeuchi}, T. \& {Lin}, D.~N.~C. 2005, \apj, 623, 482

\bibitem[{{Tanaka} \& {Tsukamoto}(2019)}]{2019MNRAS.484.1574T}
{Tanaka}, Y.~A. \& {Tsukamoto}, Y. 2019, \mnras, 484, 1574

\bibitem[{{Turner} {et~al.}(2014){Turner}, {Fromang}, {Gammie}, {Klahr},
  {Lesur}, {Wardle}, \& {Bai}}]{2014prpl.conf..411T}
{Turner}, N.~J., {Fromang}, S., {Gammie}, C., {et~al.} 2014, in Protostars and
  Planets VI, ed. H.~{Beuther}, R.~S. {Klessen}, C.~P. {Dullemond}, \&
  T.~{Henning}, 411

\bibitem[{{Tychoniec} {et~al.}(2018){Tychoniec}, {Tobin}, {Karska}, {Chand
  ler}, {Dunham}, {Harris}, {Kratter}, {Li}, {Looney}, {Melis}, {P{\'e}rez},
  {Sadavoy}, {Segura-Cox}, \& {van Dishoeck}}]{2018ApJS..238...19T}
{Tychoniec}, {\L}., {Tobin}, J.~J., {Karska}, A., {et~al.} 2018, \apjs, 238, 19

\bibitem[{{Winter} {et~al.}(2018){Winter}, {Clarke}, {Rosotti}, {Ih},
  {Facchini}, \& {Haworth}}]{2018MNRAS.478.2700W}
{Winter}, A.~J., {Clarke}, C.~J., {Rosotti}, G., {et~al.} 2018, \mnras, 478,
  2700

\bibitem[{{Yang} {et~al.}(2017){Yang}, {Johansen}, \&
  {Carrera}}]{2017A&A...606A..80Y}
{Yang}, C.-C., {Johansen}, A., \& {Carrera}, D. 2017, \aap, 606, A80

\bibitem[{{Youdin}(2010)}]{2010EAS....41..187Y}
{Youdin}, A.~N. 2010, in EAS Publications Series, ed. T.~{Montmerle},
  D.~{Ehrenreich}, \& A.~M. {Lagrange}, Vol.~41, 187--207

\bibitem[{{Youdin} \& {Shu}(2002)}]{2002ApJ...580..494Y}
{Youdin}, A.~N. \& {Shu}, F.~H. 2002, \apj, 580, 494

\bibitem[{{Zhu} {et~al.}(2010){Zhu}, {Hartmann}, \&
  {Gammie}}]{2010ApJ...713.1143Z}
{Zhu}, Z., {Hartmann}, L., \& {Gammie}, C. 2010, \apj, 713, 1143

\bibitem[{{Zhu} {et~al.}(2019){Zhu}, {Zhang}, {Jiang}, {Kataoka}, {Birnstiel},
  {Dullemond}, {Andrews}, {Huang}, {P{\'e}rez}, {Carpenter}, {Bai}, {Wilner},
  \& {Ricci}}]{2019ApJ...877L..18Z}
{Zhu}, Z., {Zhang}, S., {Jiang}, Y.-F., {et~al.} 2019, \apjl, 877, L18

\bibitem[{{Zsom} {et~al.}(2010){Zsom}, {Ormel}, {G{\"u}ttler}, {Blum}, \&
  {Dullemond}}]{2010A&A...513A..57Z}
{Zsom}, A., {Ormel}, C.~W., {G{\"u}ttler}, C., {Blum}, J., \& {Dullemond},
  C.~P. 2010, \aap, 513, A57

\end{thebibliography}

\begin{appendix}

\section{Cloud core collapse}
\label{app:collapse}


The velocity at which the collapse front expands outwards can be found from the equation of the collapse time, which \cite{2013ApJ...770...71T} found to be

\begin{align}
t(r_\mathrm{cf})=\sqrt{\dfrac{r_\mathrm{cf}^3}{2GM\left(r_\mathrm{cf}\right)} } \int_0^1 \dfrac{\mathrm{d}R}{\sqrt{\dfrac{1}{f}\ln R + \dfrac{1}{R} -1 }}, \label{eq:app:tcoll}
\end{align}
where $r_\mathrm{cf}$ is the radius of the collapse front, $G$ is the gravitational constant, $M(r_\mathrm{cf})$ is all the mass contained within the collapse front radius, and $R=r_\mathrm{cf}/r_\mathrm{cc}$ where $r_\mathrm{cc}$ is the radius of the cloud core.
Combining all the constants into one as
\begin{align}
\gamma =  \sqrt{\dfrac{1}{2 G }}\int_0^1 \dfrac{\mathrm{d}R}{\sqrt{\dfrac{1}{f}\ln R + \dfrac{1}{R} -1 }},
\end{align}
we can rewrite Eq. \eqref{eq:app:tcoll} as
\begin{align}
\int_0^{r_\mathrm{cf}} 4 \pi r^2 \rho(r) \mathrm{d}r = \dfrac{r_\mathrm{cf}^3}{t^2}\gamma^2.
\end{align}
Differentiating both sides of this equation with respect to $r_\mathrm{cf}$ gives
\begin{align}
4 \pi r_\mathrm{cf}^2 \rho(r_\mathrm{cf}) = \dfrac{3 r_\mathrm{cf}^2\ \gamma^2}{t^2} - \dfrac{2r_\mathrm{cf}^3\ \gamma^2}{t^3}\dfrac{\mathrm{d}t}{\mathrm{d}r_\mathrm{cf}}.
\end{align}
Solving this for $\mathrm{d}t/\mathrm{d}r_\mathrm{cf}$ and inverting it gives
\begin{align}
\dfrac{\mathrm{d}r_\mathrm{cf}}{\mathrm{d}t} = \dfrac{2 r_\mathrm{cf}\ \gamma^2}{3t\gamma^2 - 4\pi t^3\rho(r_\mathrm{cf})}.
\end{align}

\section{The effect of changing $\alpha_\mathrm{min}$}
\label{app:alpha}

Increasing or decreasing the minimum value, $\alpha_\mathrm{min}$, that the viscosity parameter $\alpha$ can take affects the rate of viscous evolution. To find out the effect of changing $\alpha_\mathrm{min}$ we run a case that is identical to the reference case, except that $\alpha_\mathrm{min} = 0.001$. Because the transport of gas is less efficient for lower viscosities, a lower value for $\alpha_\mathrm{min}$ will lead to a disc with a smaller radius, and hence a higher gas surface density, if the disc mass is the same. The lower viscosity will decrease the gas accretion rate compared to the reference case, but the increased gas surface density will act to increase the gas accretion rate. Therefore, the combined effect of decreasing $\alpha_\mathrm{min}$  on the gas accretion rate is not immediately clear. The matter becomes further complicated when gravitational instability is taken into account. The more compact disc from a low $\alpha_\mathrm{min}$ case can become more gravitationally unstable due to the same mass being confined to a smaller region.

\begin{figure}[]
\resizebox{\hsize}{!}{\includegraphics[width=\hsize]{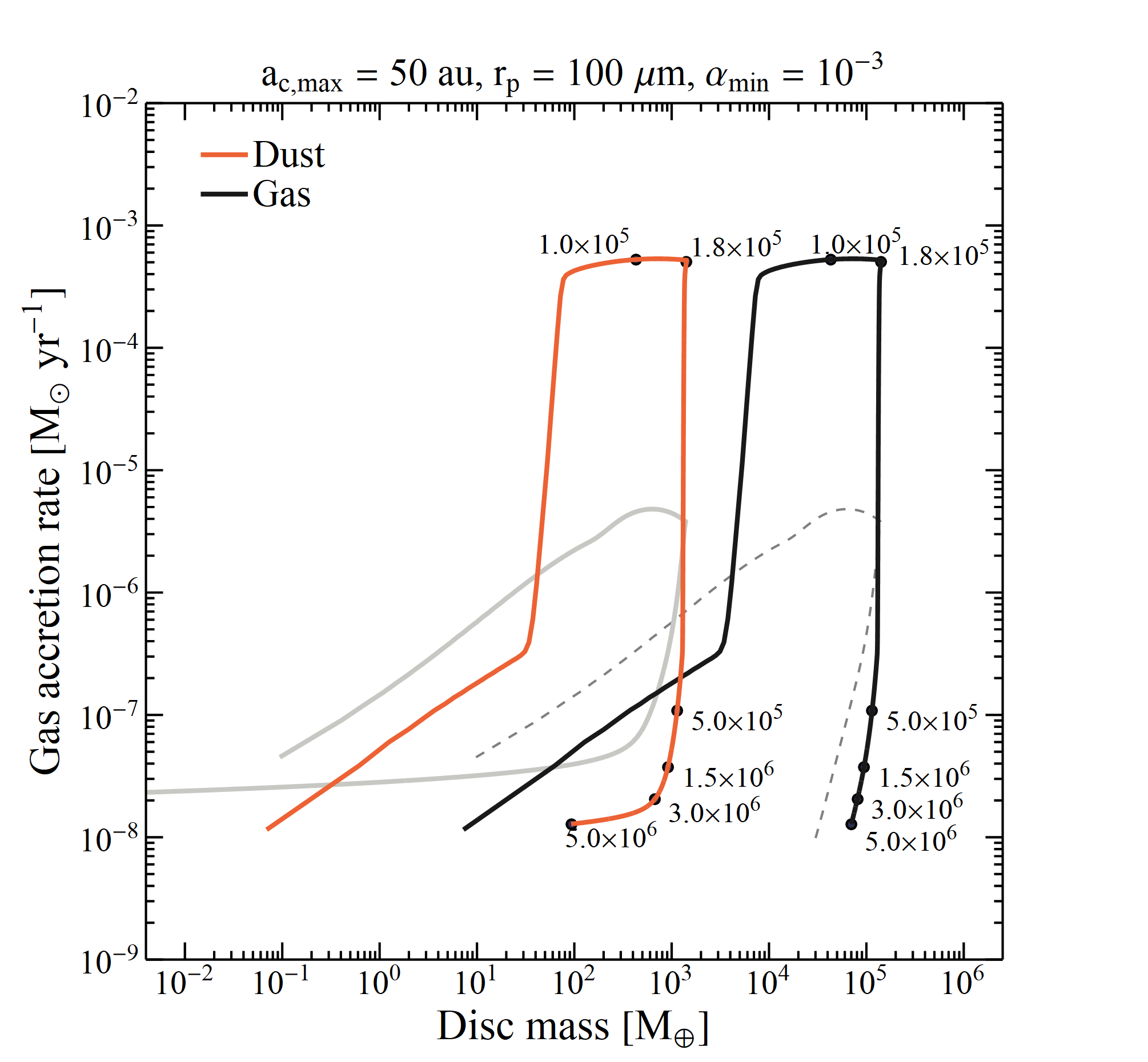} }
      \caption{ Stellar gas accretion rate as a function of disc mass using $\alpha_\mathrm{min} = 10^{-3}$. The grey lines show the reference case dust mass (solid line) and gas (dashed lines). Decreasing $\alpha_\mathrm{min}$ results in the disc becoming more gravitationally unstable. This causes the very large increase in the accretion rate in the second half of the formation phase. The accretion rate then drops rapidly until about 0.5 Myr. After this it decreases more slowly. The dust drift turn-over occurs later with lower $\alpha_\mathrm{min}$ and the dust disc drains slower. The final gas accretion rate is similar for $\alpha_\mathrm{min} = 10^{-3}$ and $\alpha_\mathrm{min} = 10^{-2}$, but the disc masses of the $\alpha_\mathrm{min} = 10^{-3}$ case are higher.
      }
         \label{Fig:app:Nom:MdotvsMdisclowa}
   \end{figure}

The result of using $\alpha_\mathrm{min} = 0.001$ is shown in Fig. \ref{Fig:app:Nom:MdotvsMdisclowa}, which shows the gas accretion rate as a function of disc mass. During the formation phase, this appears to be very different from the reference case in Fig. \ref{Fig:Nom:MdotvsMdisc}. In the very early stages, the gas accretion rate and disc mass increase in a similar way as the reference case, but at a lower accretion rate. However, once the disc becomes gravitationally unstable the accretion rate increases very quickly. This dramatic change occurs because when  $\alpha_\mathrm{min} = 0.001$ the disc does not viscously expand significantly. Therefore this disc becomes more gravitationally unstable than the reference case, and the effective $\alpha$ becomes much higher. The highest $\alpha$ value in the disc during the phase of gravitational instability is nearly two orders of magnitude higher when $\alpha_\mathrm{min} = 0.001$ compared to when $\alpha_\mathrm{min} = 0.01$. 

Once the formation phase is over, the gravitational instability phase also ends quickly and the gas accretion rate now drops very quickly at first, dropping four orders of magnitude in the first $3\times 10^5$ yr after the end of the formation phase. After this, the evolution of the accretion rate slows down. The dust drift turn-over is reached at 3.5 Myr, instead of at 1.3 Myr as for the reference case. There is also several tens of Earth masses left in the disc at 5 Myr compared to the  reference case that was completely drained in 2.5 Myr. The gas mass is about twice as high at the end of the simulation in the $\alpha_\mathrm{min} = 0.001$ compared to the reference case.

The final gas accretion rate at 5 Myr is not significantly different when using $\alpha_\mathrm{min} = 0.001$ and $\alpha_\mathrm{min} = 0.01$. Both have accretion rates of $\dot{M} \approx\! 10^{-8}$ $M_\sun$ yr$^{-1}$ at this point. 


\section{The effect of changing the dust particle size}
\label{app:dustsize}

\begin{figure}[t]
\resizebox{\hsize}{!}{\includegraphics[width=\hsize]{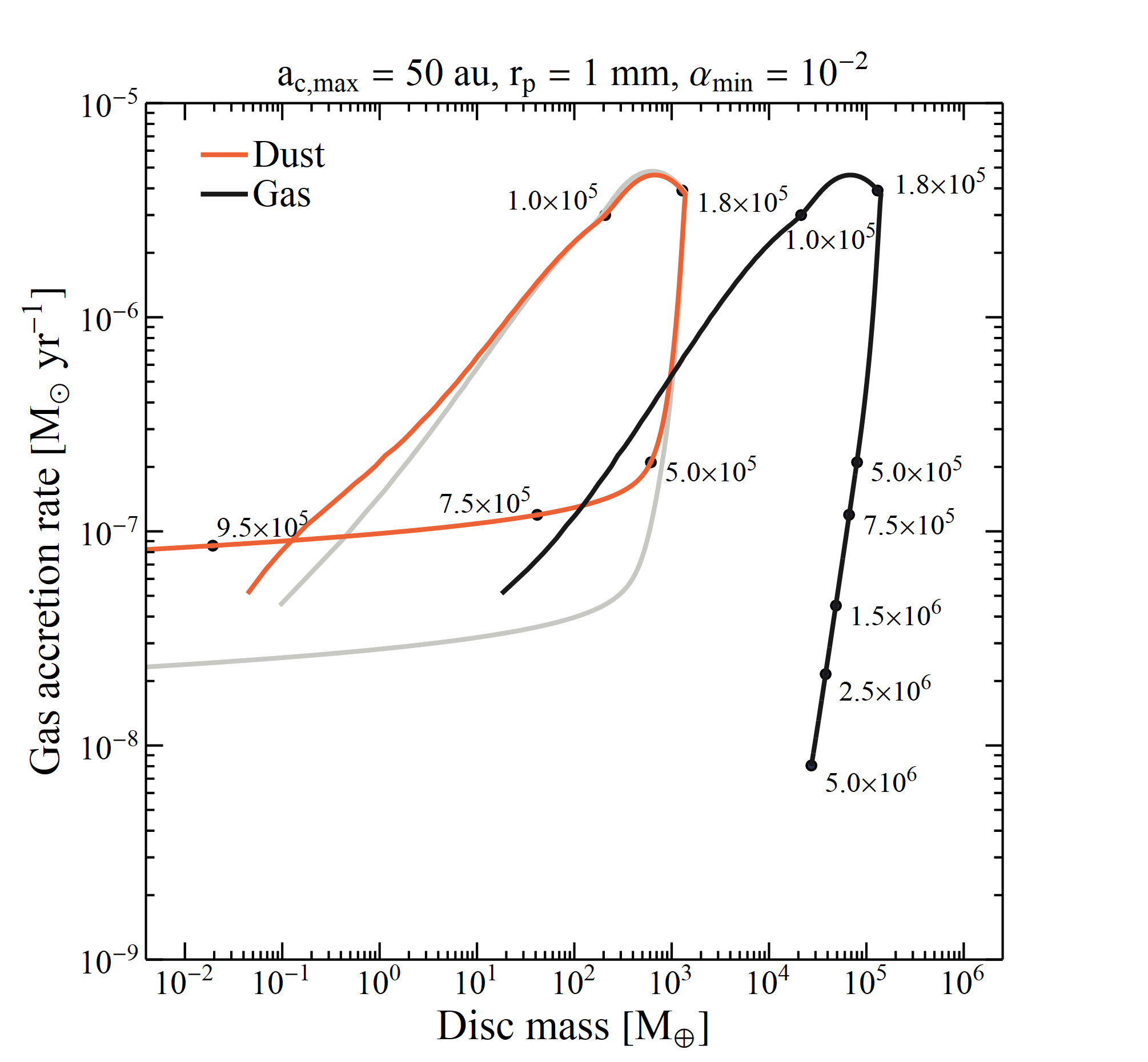} }
      \caption{Stellar gas accretion rate as a function of disc mass using a dust size of $r_\mathrm{p} = 0.1$ cm. The red curve shows the dust and the black curve shows the gas. The black dots mark points in time as labelled. The grey line shows the dust disc mass of the reference case. The rapid draining of the dust mass starts at about 0.5 Myr.  Within 1 Myr the dust disc has been drained to the point where only $0.017 \times 10^{-2}$ $M_\Earth$ of dust remains. 
         }
         \label{Fig:app:Nom:MdotvsMdiscrp1000}
   \end{figure}

In this section we analyse the effect of changing the size of the dust particles. For this we evolve a system that is identical to the reference case, but using a dust size of  $r_\mathrm{p} =1$ mm, compared to $r_\mathrm{p} =100$ $\mu$m in the  reference case.
Figure \ref{Fig:app:Nom:MdotvsMdiscrp1000} shows the stellar gas accretion rate as a function of disc mass. Increasing the dust size does not change the evolutionary path of the dust disc, but it speeds it up. Instead of draining in $\approx\! 2.5$ Myr as for the reference case, it drains in $\lesssim$ 0.8 Myr. The onset of the rapid dust drift regime is earlier for larger dust because the Stokes numbers are higher. Since the rapid drift begins earlier,  the gas accretion rate during the dust draining phase is higher.  

\begin{figure}[t]
\resizebox{\hsize}{!}{\includegraphics[width=\hsize]{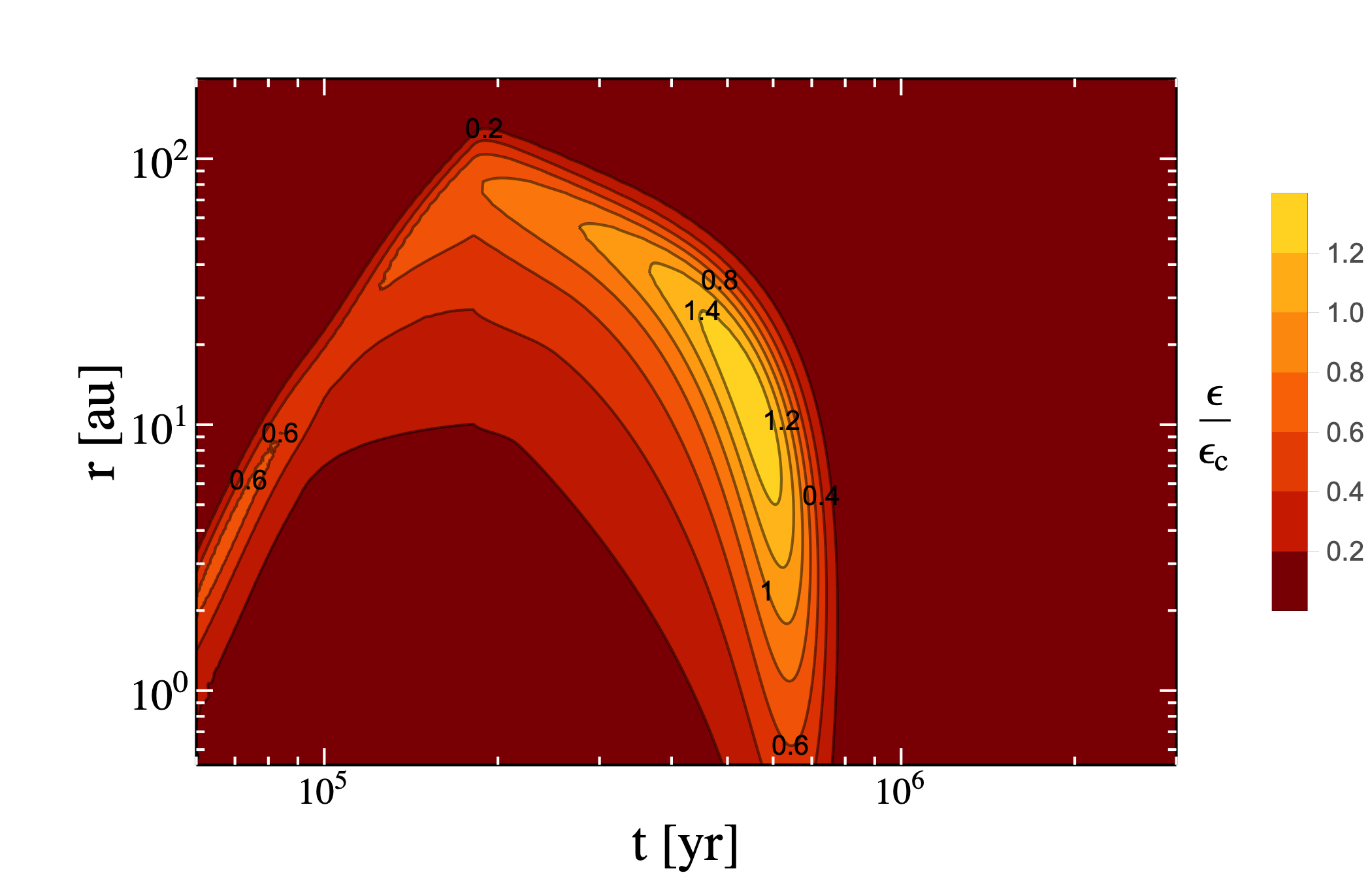} }
      \caption{Contour plot of the ratio between the critical dust-to-gas ratio and the critical dust-to-gas ratio for the streaming instability. Using dust with a size of 1 mm the critical dust-to-gas ratio passed after $\approx\! 3\times 10^5$ yr. The super-critical region remains until $t\approx\! 7 \times 10^5$ yr. }
         \label{Fig:app:CritDTGrp1000}
   \end{figure}

The dust accretion rate through the disc is also higher for 1 mm sized particles. This in turn leads to a larger increase in the dust-to-gas ratio during the phase of rapid dust drift than for the reference case. This increase also occurs earlier. The larger and earlier pile-up of dust makes this dust reach the critical dust-to-gas ratio for streaming instability earlier. This can be seen in Fig. \ref{Fig:app:CritDTGrp1000}. A super-critical region exists from $t \approx\! 0.35$ Myr to $t\approx\! 0.7$ Myr at radii between $r \approx\! 2$ au and $r \approx\! 70$ au. For the reference case the super-critical region exists rom $t \approx\! 0.8 $ Myr to $t\approx\! 1.5$ Myr at radii between $r \approx\! 8$ au and $r \approx\! 100$ au.

\section{$a_\mathrm{c}- M_\mathrm{cc}$ relation}
\label{app:BE_Rc_M}
The final centrifugal radius is given by
\begin{align}
a_\mathrm{c} = \dfrac{\Omega_0^2 r_\mathrm{cc}^4}{G M_\mathrm{cc}}, \label{eq:app:Rc}
\end{align}
where $\Omega_0$ is the rotation rate of the cloud core, $r_\mathrm{cc}$ the cloud core radius, and $M_\mathrm{cc}$ the cloud core mass.
The ratio of rotational and gravitational energy in the cloud core is given by
\begin{align}
\beta = \dfrac{E_\mathrm{rot}}{E_\mathrm{grav}} = \dfrac{\frac{1}{2}\ I \Omega_0^2}{q\dfrac{ G M_\mathrm{cc}^2}{r_\mathrm{cc}}}, \label{eq:app:beta}
\end{align}
where $I$ is the moment of inertia of the cloud core and $q$ is a parameter related to the density profile of the cloud core, e.q. $q = 3/5$ for a constant density sphere. $\beta$ is thought to be independent of the size of a cloud core \citep{1993ApJ...406..528G, 2018ApJ...865...34C}. We therefore assume $\beta$ to be the same for all cloud cores.
From equation \eqref{eq:app:beta} we find $\Omega_0^2$ as
\begin{align}
\Omega_0^2 \propto \dfrac{\beta \frac{M_\mathrm{cc}^2}{r_\mathrm{cc}}}{I}. \label{eq:app:omega^2}
\end{align}
The moment of inertia, $I$, of a sphere with varying density can be calculated as
\begin{align}
I = \dfrac{2}{3} M \dfrac{\int_0^R r^4\rho(r) \mathrm{d}r}{\int_0^R r^2\rho(r) \mathrm{d}r},
\end{align}
where $M$ is the mass of the sphere, $\rho(r)$ its density, and $R$ its radius. For a radial density profile that follows $\rho (r) \propto r^a$ where $a\neq -3$ the above relation gives
\begin{align}
I \propto M r^2.
\end{align}
Inserting this into equation \eqref{eq:app:omega^2} gives
\begin{align}
\Omega_0^2 \propto \dfrac{\beta\frac{M_\mathrm{cc}^2}{r_\mathrm{cc}}}{M_\mathrm{cc}r_\mathrm{cc}^2} = \dfrac{\beta M_\mathrm{cc}}{r_\mathrm{cc}^3}.
\end{align}
If we insert this into equation \eqref{eq:app:Rc}, using that $\beta$ is not proportional to the cloud core radius or mass we get
\begin{align}
a_\mathrm{c} \propto  r_\mathrm{cc}.
\end{align}
For a Bonnor-Ebert sphere of a given temperature the radius of the sphere is linearly proportional to the mass, giving
\begin{align}
a_\mathrm{c} \propto M_\mathrm{cc}.
\end{align}

\section{The effect of adding a metallicity spread}
\label{app:metallicity}

Here we look at how increasing or decreasing the initial dust-to-gas ratio (i.e. that of the  cloud core) affects the disc evolution. We begin by looking at two cases, one where the initial dust-to-gas ratio has been halved and another where it has been doubled, in order to understand how this affects the dust disc. We then repeat the population synthesis of a star-forming region as in the main study, but introduce a scatter in the initial dust-to-gas ratio. The amount of scatter we introduce  is based on the scatter in metallicity of nearby star-forming regions. This is quite low, and therefore including variation in the initial dust-to-gas ratio only introduces a small scatter in dust disc masses into our results.

 \begin{figure}[t]
\resizebox{\hsize}{!}{\includegraphics[width=\hsize]{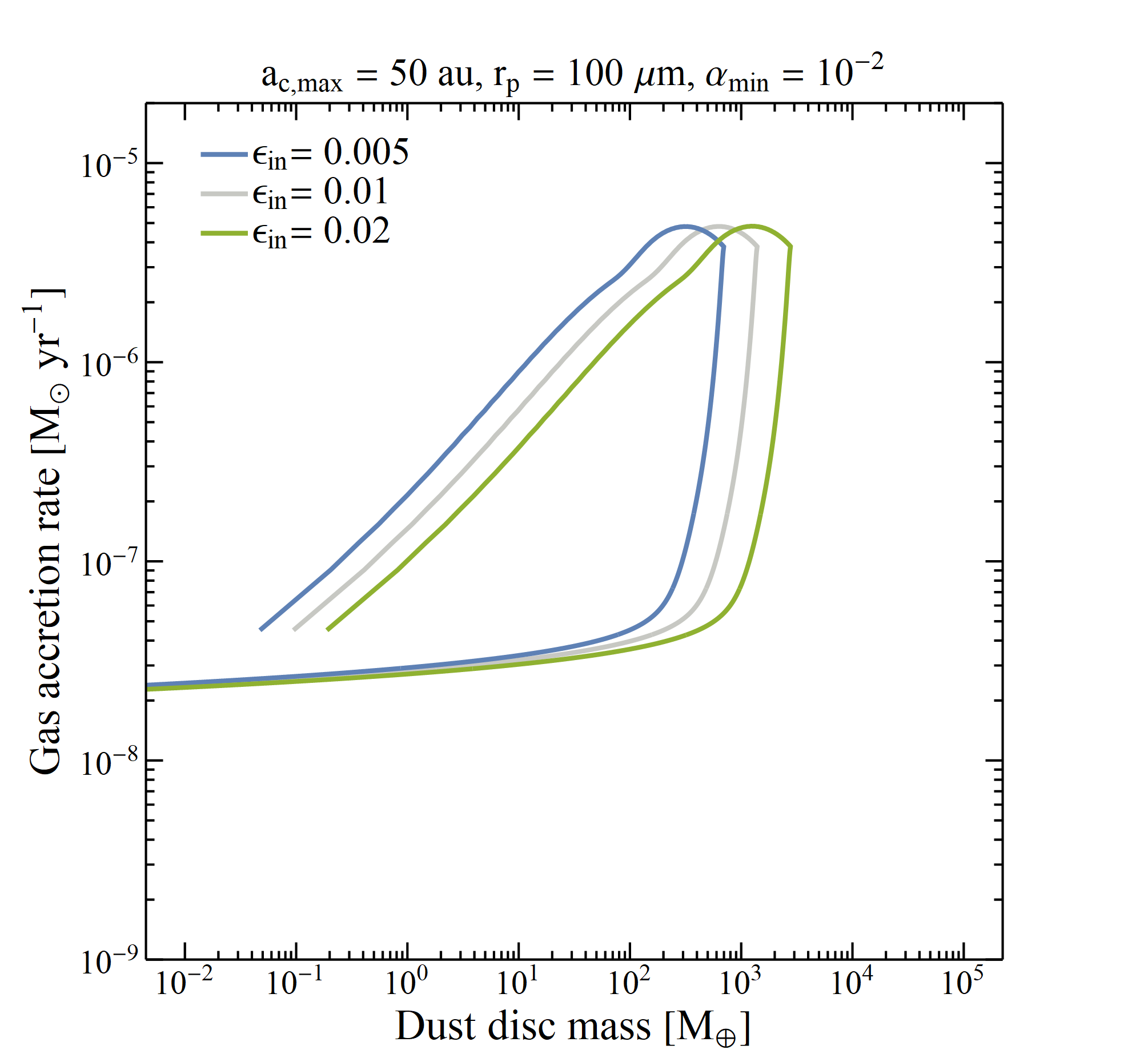} }
      \caption{Gas accretion rate as a function of dust disc mass for three different initial dust-to-gas ratio cases. The grey curve shows the reference case, the blue curve shows the case with the initial dust-to-gas ratio half of the reference case, and the green curve shows the case with double the reference case's initial dust-to-gas ratio. From this we can see that in this relation, increasing or decreasing the initial dust-to-gas ratio by some factor results in an equally large change in the dust disc mass. The dust drift turn-over point and the gas accretion rate are unaffected in our model.
         }
         \label{Fig:app:Nom:MdotvsMdiscDTGvar}
   \end{figure}

Figure \ref{Fig:app:Nom:MdotvsMdiscDTGvar} shows the gas accretion rate as a function of disc mass for the two cases with halved and doubled initial dust accretion rate in blue and green respectively. The reference case is plotted in grey. From this we can see that halving or doubling the initial dust-to-gas ratio halves or doubles the dust disc mass, respectively. Since the gas disc is unaffected by the dust in out model, the dust drift turn-over is unchanged.

An effect of increasing the initial dust-to-gas ratio is that the backreaction of dust on gas would become stronger. This effect can lead to self-induced dust traps which further increase the concentration of dust \citep{2017MNRAS.467.1984G}. However, we do not model this backreaction so we would not see such an effect in our model. The dust-growth timescale in coagulation models is also shortened for larger dust-to-gas ratios \citep{2012A&A...539A.148B}, but since we do not include dust growth in our model, it does not affect our findings. 

To a good approximation, the metallicity in the solar neighbourhood has a mean that is slightly subsolar with a small scatter $\mathrm{\left[Fe/H\right]} = -0.1  \pm 0.1\ $ dex \citep{2009MNRAS.399.1145S,2008A&A...481..199N,2019A&A...625A.120B}. This value is similar to that of stars in nearby star-forming regions where the metallicity is solar or slightly subsolar. Reported values are:  $\mathrm{\left[Fe/H\right] = -0.01 \pm 0.04}$ dex in Orion  \citep{2009A&A...501..973D},  $\mathrm{\left[Fe/H\right] = -0.01 \pm 0.05}$ dex , in Taurus \citep{2011A&A...526A.103D}, and $\mathrm{\left[Fe/H\right] = -0.08\pm 0.04}$ dex in Chameleon  \citep{2014A&A...568A...2S}.

\begin{figure}[t]
\resizebox{\hsize}{!}{\includegraphics[width=\hsize]{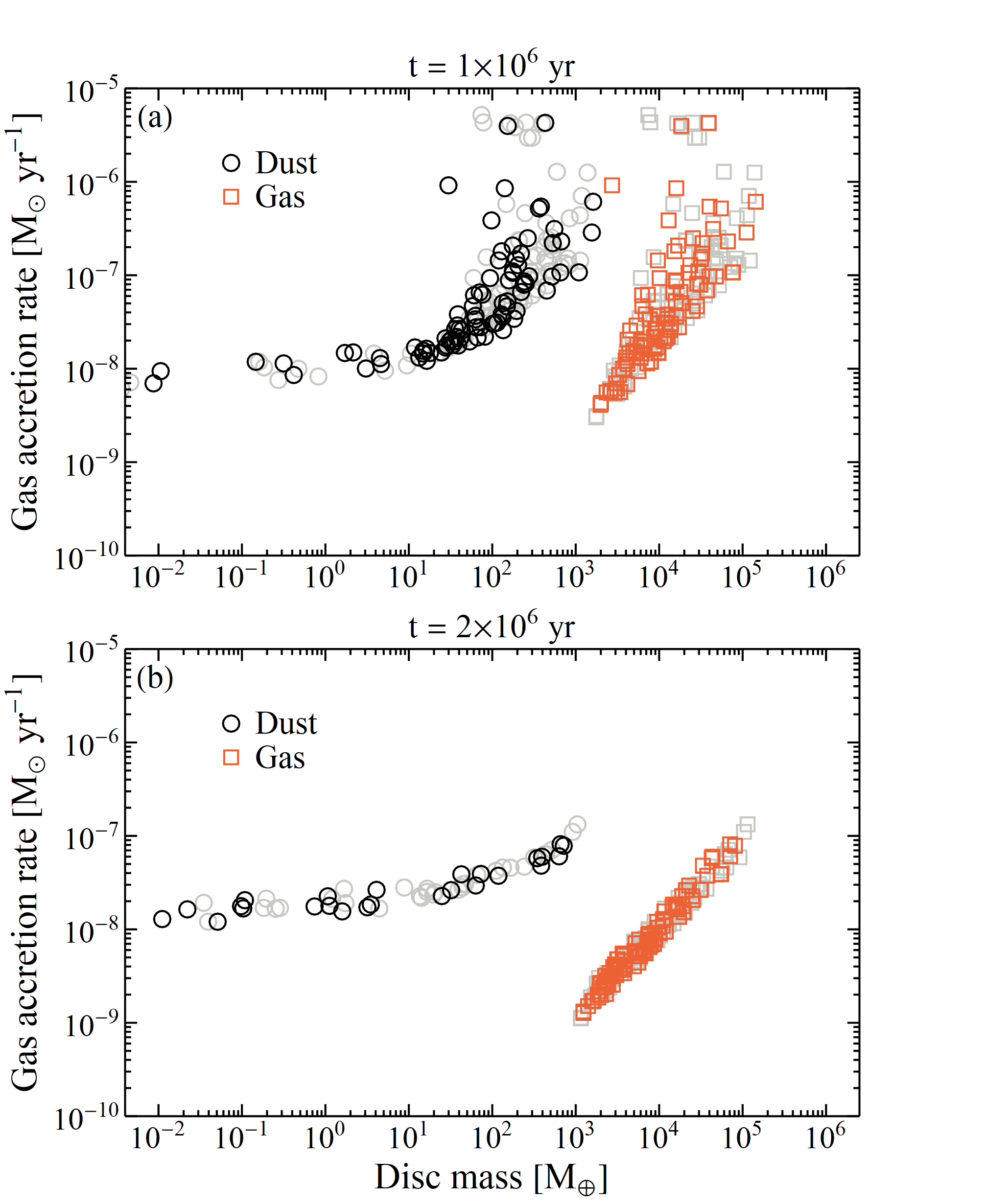} }
      \caption{Gas accretion rate as a function of disc mass for a stellar cluster where the scatter in initial dust-to-gas ratio matches that of the scatter in metallicites in the solar neighbourhood and nearby star-forming regions. Comparing this to when the initial dust to gas ratio is constant, which is shown in grey, (see also Fig. \ref{Fig:MC:MdotvsMdisc}) we can see that the two appear very similar.
   }
         \label{Fig:app:Nom:MdotvsMdiscDTGvarMc}
   \end{figure}

With this in mind, we rerun the population synthesis model but now we pick the initial dust-to-gas ratio from a normal distribution with a mean of  0.01 and a standard deviation of 0.0025 (corresponding to 0.1 dex). The result of this is shown in Fig. \ref{Fig:app:Nom:MdotvsMdiscDTGvarMc}. Comparing this to the population synthesis without a scatter in the initial dust-to-gas ratio shown in grey symbols, the results are very similar. Because the scatter of metallicities we introduce is small, most cloud cores will have an initial dust-to-gas ratio within a factor of two from the mean of 0.01. In Sect \ref{sec:res:grid:rc} and \ref{sec:res:grid:mass} we showed that variations in the centrifugal radius have a similar or larger effect on the dust disc mass for a given cloud core mass. Hence, the effect of changing the initial dust-to-gas ratio is indistinguishable from the effect of variations in the centrifugal radius when one only looks at the total mass of the dust disc. 

\end{appendix}

\end{document}